\documentclass[letterpaper,12pt]{article}
\usepackage{multibib}
\newcounter{firstbib}
\newcounter{secondbib}

\def\apj{Astrophys. J.}
\def\mnras{Mon. Not. R. Astron. Soc.}
\def\nat{Nature}

\def\araa{Ann. Rev. Astron. Astrophys.}                
\def\aap{Astron. Astrophys.}                   
\def\apjl{Astrophys. J. }                   
\def\pasj{Publ. Astron. Soc. Jap.}  
\def\araa{Ann. Rev. Astron. Astrophys.}                
\def\gca{Geochimica et Cosmochimica Acta}

\usepackage{color,hyperref}

\usepackage{epsfig}
\usepackage{ulem}
\usepackage{color, xcolor}
\usepackage{graphics,graphicx}
\usepackage{times} 
\usepackage{amssymb}
\usepackage{amsmath}
\usepackage{url}
\usepackage{multirow}
\usepackage{ctable}
\usepackage{threeparttable}
\usepackage{tocloft}

\def\msun{\hbox{$\rm\thinspace M_{\odot}$}}

\def\chandra{{\it Chandra}}
\def\xmm{{\it XMM-Newton}}

\def\ks{\hbox{$\rm\thinspace ks$}}

\def\as{\hbox{$\rm\thinspace arcsec$}}

\def\pcmsq{\hbox{$\rm\thinspace cm^{-2}$}}
\def\kmpspmpc{\hbox{$\rm\thinspace km~s^{-1}~Mpc^{-1}$}}

\def\ev{\hbox{$\rm\thinspace eV$}}
\def\kev{\hbox{$\rm\thinspace keV$}}

\def\ctsps{\hbox{$\rm\thinspace count~s^{-1}$}}

\def\ergpcmsqps{\hbox{$\rm\thinspace erg~cm^{-2}~s^{-1}$}}

\def\ergps{\hbox{$\rm\thinspace erg~s^{-1}$}}
\def\ergcmps{\hbox{$\rm\thinspace erg~cm~s^{-1}$}}

\def\msun{\hbox{$\rm\thinspace M_{\odot}$}}

\def\rg{${\it r}_{\rm g}$}

\def\nh{${\it N}_{\rm H}$}
\def\ka{$K\alpha$}

\def\chisq{{\chi^{2}}}

\def\zpo{\rm{\small ZPOWERLAW}}

\def\phabs{\rm{\small PHABS}}
\def\zphabs{\rm{\small ZPHABS}}
\def\zpcfabs{\rm{\small ZPCFABS}}
\def\diskbb{\rm{\small DISKBB}}

\def\reflionx{\rm{\small REFLIONX}}

\def\relline{\rm{\small RelLine}}
\def\pileup{\rm{\small pileup}}
\def\relconv{\rm{\small RelConv}}

\def\zxipcf{\rm{\small zxipcf}}

\def\xspec{\hbox{\small XSPEC}}

\def\grppha{\hbox{\rm{\small GRPPHA~\/}}}

\def\rmfgen{\hbox{\rm{\small RMFGEN}}}
\def\arfgen{\hbox{\rm{\small ARFGEN}}}

\def\grid25{\hbox{\rm{\small GRID25}}}

\def\epicpn{\hbox{\rm{\small EPIC-PN}}}
\def\epicmos{\hbox{\rm{\small EPIC-MOS}}}

\def\rg{${\it r}_{\rm g}$}

\def\nh{${\it N}_{\rm H}$}
\def\ka{$K\alpha$}
\def\baselinesimple{\textit{Baseline-simple}}
\def\baselinereflection{\textit{Baseline-reflection}}

\def\rx{{RX~J1131-1231}}

\title{Reflection From the  Strong Gravity Regime in a $z=0.658$ Gravitationally  Lensed-Quasar}

\author{R.~C.~Reis$^{1,2}$,  M.~T.~Reynolds$^1$, J.~M.~Miller$^1$, \& D.~J.~Walton$^3$}
\date{}

\begin{document}
\maketitle

\vspace{-5mm}
\begin{center}
 {\footnotesize
 \noindent$^1$ Department of Astronomy, University of Michigan, Ann Arbor, Michigan 48109, USA\\ 
 $^2$  Einstein Fellow\\
 $^3$  Cahill Center for Astronomy and Astrophysics, California Institute of Technology, Pasadena, California 91125, USA\\}
\vspace{10mm}

\href{http://www.nature.com/nature/journal/v507/n7491/full/nature13031.html}{\textbf{Reis et al., 2014, Nature,
    507, 207 (DOI:10.1038/nature13031)}}
\end{center}

\vspace{5mm}
\begin{abstract}\label{abstract}
\textbf{The co-evolution of a supermassive black hole with its host galaxy\cite{gebhardt00} through
  cosmic time is encoded in its spin\cite{BertiVolonteri2008, Fanidakis2011, Volonteri2012}. At
  $z>2$, supermassive black holes are thought to grow mostly by merger-driven accretion leading to
  high spin. However, it is unknown whether below $z\sim1$ these black holes continue to grow via
  coherent accretion or in a chaotic manner\cite{Kingpringle2006}, though clear differences are
  predicted\cite{Fanidakis2011,Volonteri2012} in their spin evolution. An established
  method\cite{Risaliti2013Natur} to measure the spin of black holes is via the study of relativistic
  reflection features\cite{rossfabian1993} from the inner accretion disk.  Owing to their greater
  distances, there has hitherto been no significant detection of relativistic reflection features in
  a moderate-redshift quasar. Here, we use archival data together with a new, deep observation of a
  gravitationally-lensed quasar at $z=0.658$ to rigorously detect and study reflection in this
  moderate-redshift quasar.  The level of relativistic distortion present in this reflection
  spectrum enables us to constrain the emission to originate within $\lesssim3$ gravitational radii
  from the black hole, implying a spin parameter $a=0.87^{+0.08}_{-0.15} $ at the $3\sigma$ level of
  confidence and $a>0.66$ at the $5\sigma$ level.  The high spin found here is indicative of growth
  via coherent accretion for this black hole, and suggests that black hole growth between
  $0.5\lesssim z \lesssim 1$ occurs principally by coherent rather than chaotic accretion episodes.}
\end{abstract}

\newpage
\label{introduction}
When optically-thick material, e.g. an accretion disc, is irradiated by hard X-rays, some of the
flux is reprocessed into an additional `reflected' emission component, which contains both continuum
emission and atomic features. The most prominent signature of reflection from the inner accretion
disc is typically the relativistic iron K$\alpha$ line (6.4--6.97\kev; rest frame)\cite{tanaka1995}
and the Compton reflection hump\cite{rossfabian1993} often peaking at
$20-30$\kev\ (rest-frame). However, the deep gravitational potential and strong Doppler shifts
associated with regions around black holes will also cause the forest of soft X-ray emission lines
in the $\sim0.7 -2.0$\kev\ range to be blended into a smooth emission feature, providing a natural
explanation for the ``\textit{soft-excess}" {observed} in the X-ray spectra of nearby active
galactic nuclei (AGNs)\cite{Crummy06,Waltonreisspin2013}. Indeed, both the iron line and the soft
excess can be used to provide insight into the nature of the central black hole and to measure its
spin\cite{Waltonreisspin2013}. Prior studies have revealed the presence of a \textit{soft-excess} in
$\gtrsim90\%$ of quasars at\cite{Porquet04quasars,Piconcelli05quasars} $z\lesssim1.7$, and broad
Fe-lines are also seen in $\lesssim25\%$ of these objects\cite{Porquet04quasars,jimenez05quasarsJ},
suggesting that reflection is also prevalent in these distant AGNs.  However, due to the inadequate
S/N resulting from their greater distances, the X-ray spectra of these quasars were necessarily
modeled using simple phenomenological parameterizations\cite{Piconcelli05quasars,Green2009quasar}.

\label{rxj1131}
Gravitational lensing offers a rare opportunity to study the innermost relativistic region in
distant quasars\cite{PooleyBlackburne2007,ChartasKochanek2012quasar}, by acting as a natural
telescope and magnifying the light from these sources. Quasars located between $0.5\lesssim z
\lesssim 1$ are considerably more powerful than local Seyferts and are known to be a major
contributor to the Cosmic X-ray background\cite{GilliCXB2007}, making them objects of particular
cosmological importance. 1RXS~J113151.6-123158 (hereafter \rx) is a quadruply imaged quasar at
redshift $z=0.658$ hosting a supermassive black hole ($\rm M_{BH} \sim2\times 10^8
\msun$)\cite{Sluse2012mass} gravitationally lensed by an elliptical galaxy at
$z=0.295$~\cite{Sluse2003}. X-ray and optical observations exhibited an intriguing flux-ratio
variability between the lensed images, which was subsequently revealed to be due to significant
gravitational micro-lensing by stars in the lensing
galaxy\cite{PooleyBlackburne2007,ChartasKochanek2012quasar}.

{Taking advantage} of gravitational micro-lensing techniques, augmented by substantial monitoring
with the \chandra\ X-ray observatory, a tight limit of the order of $\sim10$ gravitational radii
(${\rm r_g = GM/c^2}$) was placed\cite{DaiKochanek2010quasar} on the maximum size of the X-ray
emitting region in \rx, indicative of a highly compact\cite{reis2013corona} source of emission
($\lesssim 3~{\rm billion~km}$, or $ \lesssim 20~{\rm AU}$). The lensed nature of this quasar
provides an excellent opportunity to study the innermost regions around a black hole at a
cosmological distance (the look-back time for \rx\ is $\sim 6.1$~billion years), and to this effect
\chandra\ and \xmm\ have to-date accumulated nearly 500\ks\ of data on \rx.

\label{Resultsl}
Starting with the \chandra\ data, fits with a model consisting of a simple absorbed powerlaw
continuum, both to the data from individual lensed images (Extended~Data~Fig.~1) and to the co-added
data (Extended~Data~Fig.~2,4,5), reveal broad residual emission features both at low energies
($\lesssim2$\,keV rest frame, the ``soft-excess'') and around the iron K energies (3.5--7\,keV rest
frame), characteristic signatures of relativistic disk
reflection\cite{Waltonreisspin2013,Reynolds2013review}. To treat these residuals, we consider two
template models based on those commonly used to fit the spectra of
Seyferts\cite{FabZog09,Risaliti2013Natur} and stellar mass black hole binaries\cite{miller07review},
and which have also at times been used to model local quasars\cite{schmoll09}. The first is a simple
phenomenological combination of a power-law, a soft-thermal disk and a relativistic Fe-line
component (\textit{baseline-simple}), and the second employs a self-consistent blurred-reflection
model together with a power-law (\textit{baseline-reflection}). In addition, both models include two
neutral absorbers; the first to account for possible intrinsic absorption at the redshift of the
quasar; and the second to account for Galactic absorption.

We first statistically confirm the presence of reflection features in \rx\ using
\textit{baseline-simple}. Least-squares fits were made to all the individual \chandra\ spectra of
image-B simultaneously, allowing only the normalisations of the various components and the power-law
indices to vary (Extended~Data~Fig.~1). The thermal-component, used here as a proxy for the
soft-excess, is required at $>10\sigma$ and an F-test indicates that the addition of a relativistic
emission line to the combined \chandra\ data of image-B is significant at greater than the 99.9\%
confidence level. Tighter constraints ($>5\sigma$) on the significance of the relativistic iron line
can be obtained by co-adding all \chandra\ data to form a single, time-averaged spectrum
representative of the average behaviour of the system (Figure~2 and Extended~Data~Fig.~5).

The \xmm\ observation also shows the clear presence of a soft excess below $\sim1.2$\kev, again
significant at $>5\sigma$, and thanks to its high effective area above $\sim 5\kev$, it also
displays the presence of a hardening at high energies (Extended~Data~Fig.~6). An F-test indicates
that a break in the powerlaw at $\sim 5$\kev\ (Figure 2) is significant at the $3.6\sigma$ level of
confidence.  This hardening is consistent with the expectation of a reflection spectrum and can be
characterised with the Compton reflection hump\cite{rossfabian1993}.

\label{Spin and corona}
The unprecedented data quality for this moderate-$z$ quasar ($\sim$100,000 counts in the
0.3-8\,\kev\ energy range from each of the \chandra\ and \xmm\ datasets) enables us to apply
physically motivated{,} self-consistent models for the reflection features. We proceed by using the
\textit{baseline-reflection} model to estimate the spin parameter through a variety of analyses,
including time-resolved and time-averaged analyses of individual \chandra\ images, utilising its
superior angular resolution, and through analysis of the average spectrum obtained from all four
lensed images with \xmm. During the time-resolved analysis, the black hole spin parameter as well as
the disk inclination and emissivity profile were kept constant from epoch to epoch, while the
normalisations of the reflection and power-law components, as well as the ionisation state of the
disk and the power-law indices were allowed to vary between epochs (see online SI for further
details). In all cases, we obtain consistent estimates for the black hole spin, which imply
\rx\ hosts a rapidly rotating black hole (Extended~Data~Fig.~3 and
Extended~Data~Tables~1,2). Finally, in order to optimise the S/N and obtain the best estimate of the
spin parameter we fit the combined \chandra\ and \xmm\ data of \rx\ simultaneously with the
\textit{baseline-reflection} model and find $a=0.87^{+0.08}_{-0.15} ~{\rm Jc/GM^2}$ at the $3\sigma$
level of confidence ($a>0.66 ~{\rm Jc/GM^2}$ at the $5\sigma$ level; Figure~3).
 
The tight constraint on the spin of the black hole in this gravitationally lensed quasar represents
a robust measurement of black hole spin beyond our local universe. The compact nature of the X-ray
corona returned by the relativistic reflection model used herein confirms the prior micro-lensing
analysis\cite{PooleyBlackburne2007,ChartasKochanek2012quasar}, and hence moves the basic picture of
X-ray emission in quasars away from large X-ray coronae\cite{HaardtMaraschi91} that may blanket at
least the inner disk, and more towards a compact emitting region in the very innermost parts of the
accretion flow, consistent with models for the base of a jet\cite{FalckeMarkoff2000}.

\label{CXB}
In addition to constraining the immediate environment and spin of the black hole, the analysis
presented herein has implications for the nature of the Cosmic X-ray background. The best-fit
\textit{baseline-reflection} model to the time-averaged \chandra\ and \xmm\ spectra
(Extended~Data~Figs.~5,6) suggest that the source is at times reflection dominated, i.e., we find
the ratio of the reflected to the illuminating continuum in the \chandra\ (\xmm) data to be ${\rm
  f_{reflect}/f_{illum} } = 2.3\pm 1.2~ (0.47 \pm 0.15)$ in the 0.1 -- 10\kev\ band (local frame;
Extended~Data~Table~2). However, it must be noted that uncertainties in the size of the microlensed
regions could affect the absolute value of this ratio. Nonetheless, this analysis clearly
demonstrates the presence of a significant contribution from a reflection component to the X-ray
spectrum of this $z=0.658$ quasar.  The properties of \rx\ are
consistent\cite{Porquet04quasars,jimenez05quasarsJ,
  Piconcelli05quasars,Green2009quasar,ChartasKochanek2012quasar} with the known observational
characteristics of quasars at $0.5\lesssim z \lesssim 1$, and our results suggest that the
relativistic reflection component from the large population of unobscured quasars expected in this
epoch\cite{GilliCXB2007} could significantly contribute in the 20--30\kev\ band of the Cosmic X-ray
background.

\label{Potential uncertanties and future}
Although questions have previously been raised over whether \textit{reflection} is a unique
interpretation for the features observed in AGNs, the amassed evidence points towards this
theoretical framework\cite{FabZog09,Waltonreis2012}, and reached culmination with the launch of
NuSTAR and the strong confirmation of relativistic disk reflection from a rapidly spinning
supermassive black hole at the centre of the nearby galaxy NGC~1365\cite{Risaliti2013Natur}.
Nonetheless, there still remain possible systematic uncertainties, for example, due to the intrinsic
assumption that the disk truncates at the innermost stable circular orbit. Simulations have been
performed specifically aimed at addressing the robustness of this assumption\cite{reynoldsfabian08},
which find that emission within this radius is negligible, especially for rapidly rotating black
holes, as is the case here.
 
The ability to measure cosmological black hole spin brings with it the potential to directly study
the co-evolution of the black hole and its host galaxy\cite{gebhardt00}.  The ultimate goal is to
measure the spin in a sample of quasars as a function of redshift and to make use of the spin
distribution as a window on the history of the co-evolution of black hole and
galaxies\cite{Volonteri2012}. Our measurement of the spin in \rx~is a step along that path, and
{introduces} a possible means to begin assembling a sample of supermassive black hole spins at
moderate red-shift with current X-ray {observatories}.

\newpage
\begin{figure}
\begin{center}\label{Figure 1}
\includegraphics[width=0.6\textwidth]{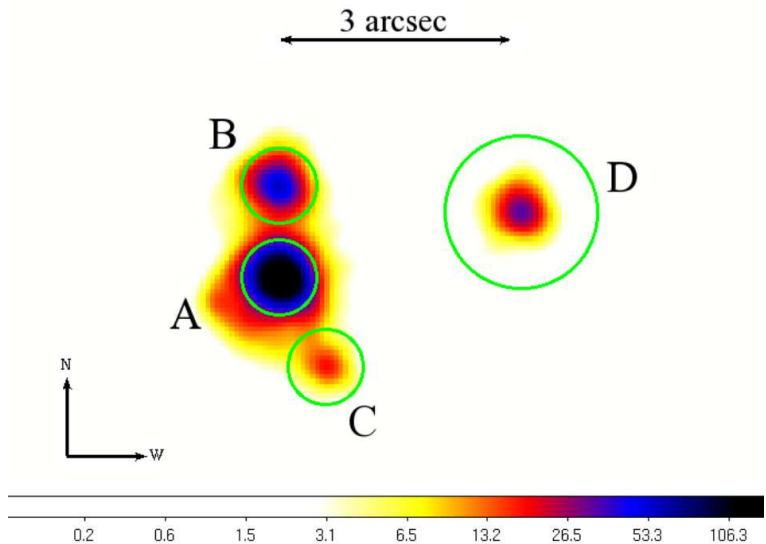}
\end{center}
\vspace*{-5mm}
\caption{{\footnotesize \textbf{\chandra\ image of RX J1131-1231}.  This representative image of a
    single epoch was made using subpixel techniques in the 0.3-8\kev\ energy range (SI) and is
    shown here smoothed with a Gaussian {($\sigma =$ 0.25'')}. The green circles show the source
    extraction regions. For images A--C, we used a radius of 0.492'', whereas the source region for
    image D was set to 0.984''. Individual source and background regions were made for all 30
    observations and spectra were extracted from the unsmoothed images. }}
\end{figure}

\newpage
\begin{figure}
\begin{center}\label{Figure 2}
\includegraphics[width=0.6\textwidth]{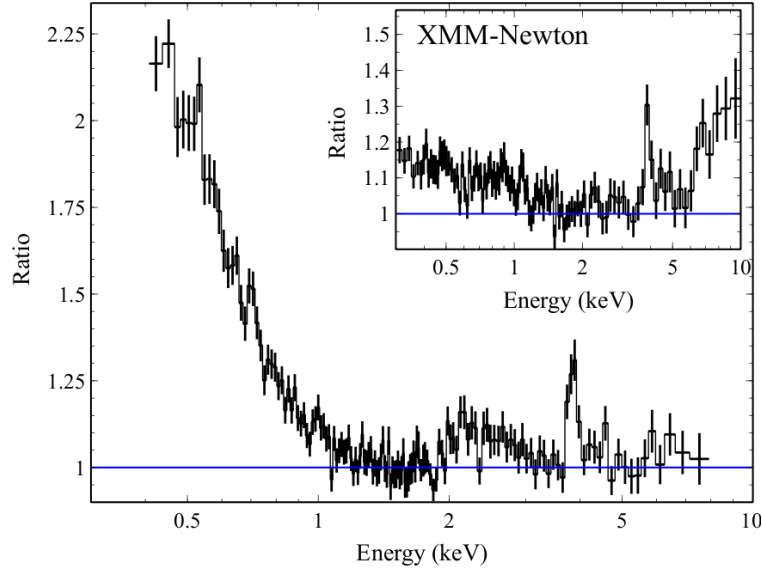}
\end{center}
\vspace*{-5mm}
\caption{{\footnotesize \textbf{Broad iron line, soft and hard excess in RX J1131-1231}. The main
    panel shows the co-added \chandra\ data over all epochs for all 4 images. The data was fit in a
    phenomenological manner with a model consisting of an absorbed power-law with an index of
    $\Gamma=1.60\pm0.04$ for the continuum, a thermal disk component with a temperature of
    $0.19\pm0.02$\kev\ to account for the soft-excess, and a broad relativistic line with energy
    constrained to lie between 6.4--6.97\kev\ (rest-frame; the \baselinesimple\ model). The ratio is
    shown after setting the normalisation of the disk, relativistic line and narrow line component
    to zero in order to better highlights these features.  The inset shows the \xmm\ data fit with a
    $\Gamma=1.83$ powerlaw. The best fit, phenomenological model for the \xmm\ data requires the
    presence of a soft excess which can again be characterised by a thermal disk component with a
    temperature of $0.22\pm0.03\kev$, a powerlaw with an index $\Gamma=1.83^{+0.07}_{-0.03}$ up to a
    break at $E_{\rm break}=5.5^{+0.5}_{-2.2}\kev$, at which point it hardens to
    $\Gamma=1.28^{+0.33}_{-0.19}$. This hardening is interpreted as the Compton reflection
    hump. Both co-added spectra shown here probe the time-averaged behaviour of \rx. Quoted errors
    refers to the 90\% confidence limit and the error bars are~1$\sigma$.  }}
\end{figure}

\newpage
\begin{figure}
\begin{center}\label{Figure 3}
\includegraphics[width=0.6\textwidth]{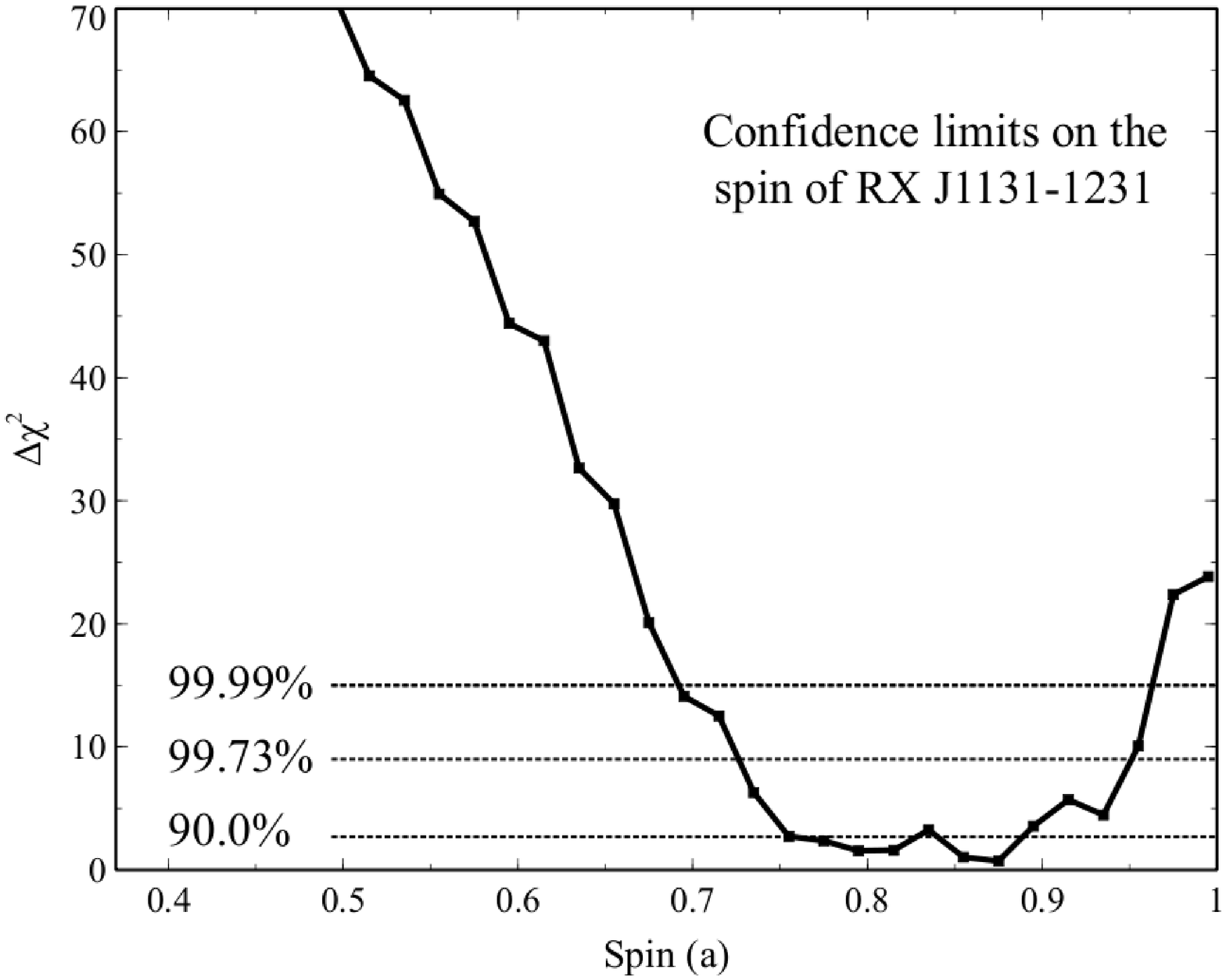}
\end{center}
\vspace*{-5mm}
\caption{{\footnotesize \textbf{Goodness-of-fit versus the spin parameter of the supermassive black
      hole in RX~J1131-1231}.  Fits were made with the spin parameter varying from 0.495 to 0.995 in
    steps of 0.02 with all all parameters of the model allowed to vary. The contour was made
    adopting a model consisting of a power-law together with a relativistic blurred reflection by an
    accretion disk, as well as two neutral absorbers; one at the redshift of the quasar and another
    local to our Galaxy (the \baselinereflection\ model). The fit was made to the co-added spectra
    from both \xmm\ and \chandra\ simultaneously, with the assumption that the spin of the black
    hole, the inclination of the accretion disk and the total hydrogen in our line of sight does not
    change between observations.  The dotted lines show the 99.99\%, 99.73\% ($3\sigma$) and 90\%
    confidence limit where it becomes clear that the supermassive black hole in RX~J1131-1231 must
    be rotating with a spin $a=0.87^{+0.08}_{-0.15} ~ {\rm Jc/GM^2}$ at the $3\sigma$ level of
    confidence.}}
\end{figure}

\newpage
\section*{Methods Summary}\label{Data and model}
We produced images for all 30 individual \chandra\ pointings (Figure~1; see Online Methods for
details), and spectra were extracted over the 0.3-8.0\kev\ energy band for each of the 4 lensed
images in each observation (all energies are quoted in the observed frame unless stated
otherwise). Previous {studies\cite{ChartasKochanek2012quasar} have demonstrated that certain lensed
  images/epochs might suffer from a moderate level of pileup\cite{pileup2010}. As such, we exclude
  spectra that displayed any significant level of pile-up in all further analysis} (see Online
Methods for details and Extended~Data~Figs.~7,8). The remaining spectra sample a period of
$\sim8$~years which allows for both a time-resolved and time-averaged analysis of \rx. We also
analyse a deep \xmm\ observation taken in July 2013, which provides an average spectrum of the four
lensed images over the 0.3--10.0\,keV energy range.\\

\noindent\textbf{Acknowledgements~} R.R. thanks the Michigan Society of Fellows and NASA for support
through the Einstein Fellowship Program, grant number PF1-120087. All authors thank the ESA
\xmm\ Project Scientist Norbert Schartel and the \xmm\ planning team for carrying out the DDT
observation. The scientific results reported in this article are based on data obtained from the
\chandra\ Data Archive.\\

\noindent\textbf{Author Contributions~} R.R. performed the data reduction and analysis of all the
data reported here. The \xmm\ data was reduced by both R.R and M.R. The pileup study was carried out
by R.R, J.M and M.R. The text was composed, and the paper synthesised by R.R, with help from D.W and
M.R. The smoothed subpixel images were made by R.R and M.R.  All authors discussed the results and
commented on the manuscript.\\

\noindent\textbf{Author Information~} Reprints and permissions information is available at
www.nature.com/reprints. The authors declare that they have no competing financial
interests. Correspondence and requests for materials should be addressed to R.~C.~Reis.~(email:
rdosreis@umich.edu).\\

\vspace{1cm}
\setcounter{firstbib}{0}

\newpage
\begin{center}
{\Large Online Methods}\label{METHODS}
\end{center}

\section{Data Reduction}
\subsection{Chandra:}\label{dataReduction}
\addcontentsline{toc}{subsection}{A: \chandra}

All publicly available data on \rx\ was downloaded from the \chandra\ archive.  As of March 13,
2013, this totalled 30 individual pointings and 347.4\ks\ of exposure, during a baseline of nearly 8
years starting on April 12, 2004 (ObsID 4814) and ending on November 9, 2011 (ObsID 12834). We refer
the reader to\cite{size1104, DaiKochanek2010quasar, ChartasKochanek2012quasar} for details of the
observations. We note that the work presented herein includes one extra epoch that was not used in
the work of \cite{ChartasKochanek2012quasar}. This further observation (ObsID 12834) added
13.6\ks\ to their sample.  Starting from the raw files, we reprocessed all data using the standard
tools available\cite{ciao} in CIAO~4.5 and the latest version of the relevant calibration files,
using the \textit{chandra\_repro} script.

Subpixel images were created for each observation and one such image is shown in Figure 1 of the
main manuscript (observation made on November 28, 2009; Sequence Number 702126; Obs ID number
11540). Sub-pixel event repositioning and binning techniques are now available\cite{Tsunemi01},
which improve the spatial resolution of \chandra\ beyond the limit imposed by the ACIS pixel size
($0.492''\times0.492''$). This algorithm, EDSER, is now implemented in CIAO and the standard
\chandra\ pipeline. Rebinning the raw data to 1/8 the native pixel size takes advantage of the
telescope dithering to provide resolution $\sim0.25''$. The EDSER algorithm now makes ACIS-S the
highest resolution imager onboard the \chandra\ X-ray Observatory. For an example,
see\cite{Wang2011subpix} for a detailed imaging study of the nuclear region of NGC 4151.

Spectra were obtained from circular regions of radius 0.492\as\ centred on Images-A, B and C as
shown in Figure 1 and from a circular region of radius 0.984\as\ for the relatively isolated
Image-D. Background spectra were taken from regions of the same size as the source located 4''
away. In the case of images-B and C, the backgrounds were taken from regions north and south of the
sources, respectively. Due to the high flux present in image-A, the presence of a read-out streak
was clear in some observations. In those cases, the background for image-A was taken from a region
centred on the read-out streak 4'' to the east of the source. When the readout streak intercepted
image-D, the background for the latter was taken from a region also centred on the readout streak
4'' to the NW. In all other cases, the background for image-D was taken from a region to the west of
the source.  Source and background spectra were then produced using \textit{specextract} in a
standard manner with the \textit{correctpsf} parameter set to ``yes".

We produced 4 spectra representing images-A, B, C and D for each of the 30 epochs as well as 4
corresponding background for each epoch. All spectra were fit in the 0.3-8.0\kev\ energy range
(observed frame) unless otherwise noted, and the data were binned using \grppha\ to have a minimum
of 20 counts per bin to assure the validity of $\chisq$ fitting statistics.

Some observations are known\cite{ChartasKochanek2012quasar} to suffer from the effects of
pile-up\cite{pile_estimate,pileup2010}, we explore in \S~2 the effect this might have on the
results.\\

\subsection{XMM-Newton:}
\addcontentsline{toc}{subsection}{B: \xmm}
We were awarded a 93\ks\ observation with \xmm\ via the Director's Discretionary Time program (Obs
ID: 0727960301) starting on 2013-07-06. The observation was made with both the \epicpn\ and
\epicmos\ in the small window mode to ensure a spectrum free of pile-up.  The level 1 data files
were reduced in the standard manner using the SAS v11.0.1 suite, following the guidelines outlined
in the \xmm\ analysis threads which can be found at
\href{http://xmm.esac.esa.int/sas/current/documentation/threads/}{(http://xmm.esac.esa.int/sas/current/documentation/threads/}. Some
background flaring was present in the last $\sim 30$\ks\ of the observation and this was removed by
ignoring periods when the 10--12\kev\ (PATTERN$==0$) count rate exceeded 0.4 ct/s, again following
standard procedures. Spectra were extracted from a 30\as\ radius region centered on the source with
the background extracted from a source free 52\as\ radius region elsewhere on the same chip. The
spectra were extracted after excluding bad pixels and pixels at the edge of the detector, and we
only consider single and double patterned events. Response files were created in the standard manner
using \rmfgen\ and \arfgen. Finally, the spectrum was rebinned with the tool \grppha\ to have at
least 25 counts per channel and was modelled over the 0.3-10\kev\ range.  We also have experimented
with \textit{specgroup} and grouped the data to a minimum S/N of 3, 5 and 10. We find that in all
cases, the results are statistically indistinguishable from the “group min 25” command in \grppha.

As the observation was taken in the small window mode with a live time of 71\% the final good
exposure, after the exclusion of the background flares identified with \epicpn, was 59.3\ks. The
observed \epicpn\ flux of $\sim1.41\ctsps$ is $\sim20$ times below the levels where pile-up is
expected to occur for this observational mode (see \xmm\ documentations at
\\ \href{http://xmm.esac.esa.int/external/xmm\_user\_support/documentation/uhb/epicmode.html}{http://xmm.esac.esa.int/external/xmm\_user\_support/documentation/uhb/epicmode.html}.

The \epicpn\ camera has the highest collecting area across the full 0.3--10.0\kev\ band, and it is
also the best calibrated camera for spectral fitting, therefore we have chosen to base our analysis
on the spectrum obtained with this detector. However, we note that similar conclusions are also
found with the \epicmos. The \epicpn\ \xmm\ spectrum is explored fully in the online SI.\\

\newpage
\section{Photon Pile-up and \chandra}
\subsection{Defining photon pile-up and the use of pile-up fraction:}
\addcontentsline{toc}{subsection}{A: Defining photon pile-up and the use of pile-up fraction}

Pile-up occurs at high fluxes when multiple photons impact a detector pixel at approximately the
same time (i.e., less than the detector frame time), and are recorded as a single event of higher
energy.  A thorough description of the impact of pile-up in the X-ray spectra of bright sources
observed by \chandra\ can be found in the \chandra\ ABC guide to pileup
at\\ \href{http://cxc.harvard.edu/ciao/download/doc/pileup_abc.pdf}{http://cxc.harvard.edu/ciao/download/doc/pileup\_abc.pdf}\\ In
the document described above, the concept of ``pile-up fraction" is used as a diagnostic of the
level of pile-up experienced in a given observation.  The pile-up fraction is expressed as a
function of ``Detected Counts per Frame" in Figure~3 of that work, where it is clear that even for
the unrealistic case where all piled events are retained as `good events' (i.e. a grade migration
parameter of $\alpha = 1$), the pile-up fraction remains below 10\% at $\sim0.18$ detected counts
per frame.  For the suggested grade migration parameter of $\alpha = 0.5$, the 10\% pile up level is
closer to $\sim0.3$ detected counts per frame.

We shown in Extended~Data~Fig.~7, the detected counts per frame over the entire 0.1-12\kev\ energy
band for all images during the 30 epochs presented here. This quantity should not be confused with
the count rate of a given observation, as it depends on the ACIS-frame time employed during the
observation. Nor is it to be confused with the unknown \textit{incident} counts per detector frame.
For a summary of the frame-time (we use frame-time to mean the sum of the static exposure time for a
frame plus the charge transfer time), and exposure employed in the various observations presented
here, see Table 1 of \cite{ChartasKochanek2012quasar}. Note that the ACIS frame time for ObsId 12833
(epoch~29) is in fact 0.441 seconds and not 0.741 seconds as reported in
ref\cite{ChartasKochanek2012quasar}. The frame time for ObsId 12834 (epoch~30) is also 0.441
seconds.

In generating this figure, we have applied a point spread function (PSF) correction to account for
the 0.492\as\ aperture used for Images-A,B,C, and 0.984\as\ for Image-D. This correction was applied
using the \textit{arfcorr} tool detailed in
\href{http://cxc.harvard.edu/ciao/ahelp/arfcorr.html}{http://cxc.harvard.edu/ciao/ahelp/arfcorr.html}. \textit{arfcorr}
estimate the fraction of the source's count lying in the extraction region at each energy, and
creates a copy of the ARF with the energy-dependent correction applied. The correction was
calculated in 4 different energy bins and the value for the counts enclosed in the 1.0--5.3\kev\ bin
was used to scale the detected counts per frame as shown in Extended~Data~Fig.~7. For all 30 epochs,
the PSF correction caused an average increase in the true detected count rate by a factor of 1.69,
1.71, 1.69 and 1.21 for images A, B,C and D, respectively; The range for these factors are
1.45--2.09, 1.44--2.08, 1.45--2.05 and 1.17--1.24 for the four images respectively over all epochs.

We also show in Extended~Data~Fig.~7 the expected ``detected counts per frame" for a a pile-up
fraction of 10\% assuming both an extreme grade migration parameter of $\alpha = 1$ (dotted) and the
more physical value of $\alpha = 0.5$ (solid), as obtained directly from Figure~3 of the
\chandra\ ABC guide to pile-up described above.

It is clear that a fraction of Image-A observations suffer from pile-up at $>10\%$ level independent
of the $\alpha$ parameter used. However, Image-B is only significantly affected by pile-up in the
first epoch. We note that this first observation is the only one having a frame-time close to the
nominal value of 3.241 seconds. All subsequent observations were performed with frame times $<1$
second which is one of the standard procedures recommended to minimise pile-up.

Here it is important to note that in observations where pile-up is important, \textit{robust}
estimates of the pileup fraction are not possible given its complex nature, and whether any given
pileup fraction should be considered ``significant" depends upon the scientific questions being
addressed by the data. We show throughout this work that the results are not influenced by pileup.\\

\subsection{Estimates of photon pile-up via grade ratios:}
\addcontentsline{toc}{subsection}{B: Estimates of photon pile-up via grade ratios}

In order to further investigate the effect of pile-up, radial plots of the ratio of \textit{ASCA}
bad grades to good grades were created, i.e., g$_{157}$/g$_{02346}$ vs r. A ratio less than 0.1
indicates that pile-up is less than 10\% (e.g., ref\cite{Russell2010pileup}). We focus on two
observations, the very first observation in which Image-B is highly piled-up (epoch 1) and the
highlighted epoch where Image-A is piled-up (epoch 23). In both observations the ratio of bad to
good events is in excess of 0.1, where we find g$_{157}$/g$_{02346} \sim$ 0.14 and $\sim$ 0.2 inside
the inner arcsecond centred on Images B and A in epochs 1 and 23 respectively. This demonstrates
that pile-up is a concern in both observations in agreement with Extended~Data~Fig.~7, which shows
these images having a pile-up fraction between $\sim25-30$\%. In addition, we note that a
readout-streak is present in epoch 23, facilitating a determination of the true spectral shape. The
events in the readout streak are those detected in the 40$\mu$s during which the ACIS detector is
being read-out and as such these events are not piled-up. Following standard procedures:
\href{http://cxc.harvard.edu/ciao/threads/streakextract/}{http://cxc.harvard.edu/ciao/threads/streakextract/}\\ the
events in the readout streak were extracted, revealing a \textit{raw} count rate of ${\rm \sim
  1~cts~per~frame}$, consistent with expectations from Extended~Data~Fig.~7.  The spectrum of
Image-A in epoch-23 (the image with the largest pile-up fraction of all epochs) was extracted from
the readout streak. Fitting this spectrum with a simple power-law model revealed a spectral index
$\Gamma \sim 1.8$. For comparison, an extraction of the piled-up Image-A spectrum returns a spectral
index of $\Gamma \sim 1.6$.

In the work that follows, we ignore all data with a pile-up fraction greater than $\sim 10\%$
assuming grade migration parameter of $\alpha=0.5$. We also ignore the first epoch of Image-C and
epoch-8 of Image-B. The results presented both in the paragraphs above and in the section that
follows clearly confirm that the disk line properties of this moderate-\textit{z} quasar are clearly
unaffected at this level. The consistency between the results obtained with \chandra\ and \xmm\ also
attest to this robustness. This cut resulted in a combined exposure of 1.13~Msec from the remaining
16, 27, 29 and 30 spectra of Images-A, B, C and D, respectively. To begin, we will consider an
observation from a single epoch (Epoch-23 mentioned above) and examine it in detail.\\

\subsection{Similarities of brighter and dimmer images:}
\addcontentsline{toc}{subsection}{C:  Similarities of brighter and dimmer images}

We show in Extended~Data~Fig.~8 (a) the spectra of Images-B, C and D for the 27.5\ks\ observation
that took place on 2009 November 28 (ObsID 11540; epoch 23). This observation is highlighted in
Extended~Data~Fig.~7 and is representative of the brightest \chandra\ epochs used in this work. It
is also the observation in which Image-A is at its brightest and therefore presents the largest
chance of cross contamination.

As a first attempt at modelling the observed spectra of epoch~23 with a simple power-law, we use the
pile-up model of ref\cite{pile_estimate} (\pileup\ in \xspec) which is explicitly designed for use
on the brightest point sources observed by \chandra. We set the frame-time parameter to
0.741~seconds and use the default values for the remaining parameters. The grade migration parameter
$\alpha$ is allowed to be free for Image-B, and is set to 0 for Images-C and D (i.e. C and D are
assumed to not suffer from any amount of pile-up). A single power-law, even after accounting for
possible pile-up affects in Image-B, does not provide a satisfactory description of the data with
$\chisq/\nu=495.1/411 =1.2$ and broad emission features are seen below $\sim$1 keV and at $\sim$3
keV, as shown in Extended~Data~Fig.~8. Adding a \diskbb\ to phenomenologically account for the soft
excess improves the fit dramatically with $\Delta\chisq/\Delta\nu =-33.1/-2$, and the remaining
excess at $\sim$3 keV is successfully modelled with the relativistic line model ``\relline ". It is
clear from Extended~Data~Fig.~8, that both the bright Image-B as well as the dimmer Images-C and D
all display similar residuals to a power-law. We therefore proceed by fitting these residuals with
the \textit{baseline-simple} model described below.

We fit the spectra of bright Image-B as well as the dimmer Images-C and D, with an absorbed
power-law together with a relativistic line (\relline\ model\cite{relconv} in \xspec) and disk
black-body component (\diskbb\ model in \xspec) to mimic the soft-excess. The relativistic line is
constrained to lie between 6.4--6.97\kev\ (rest frame).  The total model is also affected by two
neutral absorbers (\phabs\ model in \xspec); the first is to account for possible intrinsic
absorption at the redshift of the quasar\cite{Sluse2003} ($z=0.658$); and the second to account for
Galactic absorption with a column density of\cite{Dickey90} \nh$=3.6\times10^{20}\pcmsq$.  In
\xspec\ terminology this model reads $$\zphabs\times\phabs\times(\zpo+\diskbb+\relline).$$ We used
the standard \textsc{bcmc} cross-sections\cite{balucinska} and \textsc{angr}
abundances\cite{abundances} throughout this work but note that the results are not sensitive to
different abundances or different models for the neutral absorption. The three spectra are fit
simultaneously with a normalisation constant allowed to vary between them. Given that the three
spectra describe the source at different times\cite{Tewes2012}, we also allow the power-law indices
to vary. This model provides a good phenomenological description of the data
($\chisq/\nu=457.7/405=1.13$) as can be seen from the data-model ratio plot in Extended~Data~Fig.~8
(a; middle panel).

We show in panel b of Extended~Data~Fig.~8, the ratio after the removal of the soft-component and
the relativistic line (top) as well as to a simple power-law after re-fitting the data
(bottom). This simple power-law model is not a good representation of the data ($\chisq/\nu
=516.8/412=1.25$). It is clear that \textit{ all three images in this epoch display similar features
  when modelled with a simple power-law}. The fact that Images-B and C yield nearly identical
power-law indices ($\Gamma_{\rm B} = 1.61^{+0.10}_{-0.12}$ and $\Gamma_{\rm C} =
1.61^{+0.12}_{-0.08}$, respectively) -- despite the factor of $\sim3$ difference in their observed
fluxes -- would suggest that they are not suffering from the effects of pile-up -- which increases
with flux -- in accordance with the low pile-up fraction found in Extended~Data~Fig.~7. However,
their proximity to Image-A could potentially result in flux contamination. Image-D on the other hand
will not, under any scenario, suffer from cross-contamination from Image-A and the residuals to a
power-law again look remarkably similar to the residuals present in Images-B and C
(Extended~Data~Fig.~8; b), further arguing for the physical origin of these residuals. Nevertheless,
we proceed by exploring whether these residuals could be an artificial effect due pile-up or
contamination from the brighter image-A, under the flux levels characteristic of this bright epoch.\\

\subsection{MARX simulations:} 
\addcontentsline{toc}{subsection}{D: MARX simulations} In order to undertake detailed ray-tracing
simulations of the 4 images reported in this work, we make use of the latest version of the MARX
suite of programs (MARX~5.0.0) which can be found, together with the user's manual,
at\\ \href{http://space.mit.edu/cxc/marx/index.html}{http://space.mit.edu/cxc/marx/index.html}.

The methodology employed here follows directly from that recommended in the \chandra\ ABC guide to
pile-up, and involves simulating a simple $\Gamma=1.8$ power-law at the exact coordinates of the 4
images.  For each image we obtained the total 0.3-10\kev\ flux observed during epoch 23, and
simulated an observation with a total exposure of 28\ks\ with MARX, using these fluxes as input. The
{\sc marxpileup} program was then run on the simulated event file with the frametime set to
0.7~seconds and the grade migration parameter $\alpha=0.9$. This is higher than the default value of
$\alpha=0.5$, which serves to exaggerate the effects of pile-up.  We note that Image-A is being
simulated here only to asses its impact on the other images, and also that this epoch is the one
where Image-A is at its brightest (see Extended~Data~Fig.~2).

The simulated event file was reprocessed, and spectra were extracted and grouped for each image in
the exact same manner as the real data.  A power-law fit to the simulated spectrum of Image-B yields
an index of $1.78\pm0.04$ at the 90\% confidence level, in perfect agreement with the input
spectrum. No further components were needed on top of the power-law.  Again we emphasise that this
epoch is not representative of the remainder of the data and in fact represents the largest possible
level of contamination experienced over all epochs. We have repeated the simulations with Images-B
and C having a $\Gamma=1.6$ power-law, whereas we have set Images-A and D to have a different value
of $\Gamma=1.8$.  The highly piled-up spectrum of Image-A returns $\Gamma\sim1.45$ due to the
particularly high value of the grade migration parameter $\alpha$ used in our simulation.
Nonetheless, even in this extreme example, we again recover the input, featureless spectrum for
Image-B and the index is found to be $1.59\pm0.04$, in excellent agreement with the input value.

It is clear that even in our brightest observation (epoch 23; see Extended~Data~Fig.~7), an intrinsic
power-law spectrum would not be modified by either pile-up or cross contamination in such a way as to
explain the excess below $\approx1$\kev\ and between $\approx2-4$\kev\ seen in Extended~Data~Fig.~8.\\

\subsection{Summary of photon pile-up and Chandra:} 
\addcontentsline{toc}{subsection}{E: Summary of photon pile-up and Chandra} Here, we have presented
a detailed study on the effect of pile-up on our data. We have shown that there are instances when
the data suffers from moderate levels of pile-up ($>10\%$; Extended~Data~Fig.~7) and went on to
remove these data from our analyses. By using the three images (Images-B,C and D) from the brightest
epoch as a conservative example, we have shown that the residuals to a powerlaw remain present in
all cases (Extended~Data~Fig.~8), despite the different flux levels. This is contrary to the
expected behaviour if the data were suffering from significant pile-up, and strongly supports a
physical origin for these features, interpreted in this work as a soft-excess and a broad iron line.

As a further step in assessing any possible contribution pile-up may have in producing the features
observed in the \chandra\ data, we performed detailed MARX simulations, again conservatively based
on the brightest epoch, i.e. the most susceptible to pile-up. To our knowledge, MARX simulations
similar to those presented here provide the best estimate in characterising the complex effect of
pile-up, grade migration and cross-contamination. These simulations confirm that, even during this
epoch, the data from images B and C should not be contaminated by the piled-up data in image A. We
also note here that the \xmm\ observation (which is over an order of magnitude below the pileup
threshold) detailed in the Online-SI also displays similar features to those seen in the
\chandra\ data, including a highly significant soft excess and Compton hump, and returns a
consistent estimate for the black hole spin.\\

\vspace{1cm}
\setcounter{firstbib}{30}

\newpage
\begin{center}
{\Large Supplementary Information}\label{SI}
\end{center}
\setcounter{section}{0}

\section{Summary}
Here, we summarise briefly the key points demonstrated in the supplementary
information. Full details regarding the analysis performed for the quadruply
imaged quasar 1RXS~J113151.6-123158 (hereafter \rx) are given in the following sections.

\subsection{A: The spin measurements with \chandra\ are found to be consistent for a variety of analysis techniques:}
We demonstrate the presence of residuals to the standard powerlaw AGN continuum consistent with the
soft excess commonly observed in local, unobscured Seyfert galaxies, and also with a
relativistically broadened iron emission line, from which the spin of the black hole can be
constrained. We do so first using a phenomenological model including a relativistic line profile,
and then with a fully physically self-consistent reflection model, comparing the results obtained
with a time-averaged and time-resolved analyses of the data from individual \chandra\ images. The
spin constraints obtained with these various analyses are all found to be consistent, implying a
rapidly rotating black hole.\\

\subsection{B: Spin determination is consistent for both \xmm\ and \chandra:}
Having demonstrated the consistency of the results obtained with the individual \chandra\ images, we
then constrain the spin of \rx\ using all the selected \chandra\ data simultaneously with the
self-consistent reflection model, and obtain $a=0.90^{+0.07}_{-0.15}$ at 3$\sigma$ confidence. We
also constrain the spin in the same manner with an independent \xmm\ observation, obtaining
$a=0.64^{+0.33}_{-0.14}$ (again, 3$\sigma$ confidence), fully consistent with the
\chandra\ constraint. Finally, modeling both the \chandra\ and \xmm\ datasets simultaneously in
order to obtain the most robust measurement, we constrain the spin of \rx\ to be
$a=0.87^{+0.08}_{-0.15}$ at the 3$\sigma$ level of confidence.\\

\subsection{C: The spin measurements are robust against absorption:}
Lastly, we consider whether there is any evidence for absorption by partially ionised material,
often seen in local Seyferts and other quasars, in the spectrum of \rx, and investigate any effect
this might have on the spin constraint obtained. Through phenomenological modelling, we show that
although ionised absorption could plausibly reproduce the soft excess, a relativistic iron emission
line is still required, and a high spin is again obtained. Furthermore, when considering the
self-consistent reflection model, which includes the soft emission lines that naturally accompany
the iron emission, the addition of ionised absorption to the model does not improve the fit, and the
spin constraint obtained again remains unchanged.\\

\newpage 
\section{Background and Representative Values for 1RXS~J113151.6-123158 (\rx)} \label{rxj1131}
\begin{itemize}
\item Mass in the range of $\sim 8\times10^{7}~\msun$ (via H$\beta$ line\cite{Sluse2012mass}) to
  $M_{BH} \sim 2\times10^{8}~\msun$ (via MgII line).  However, the value for the dimensionless spin
  parameter presented in this work does not depend on the mass of the black hole.

\item  Quasar\cite{Sluse2003} at $z = 0.658$, lensing galaxy at at $z = 0.295$.

\item Intrinsic (non magnified) bolometric luminosity $\rm log_{10}L_{Bol} \approx 45 \ergps$
  assuming a magnification factor of 11.6 and a bolometric correction of 9.6 for image~B. See
  \cite{Sluse2012mass} for details.
\end{itemize}

\noindent{The time-averaged, unabsorbed fluxes (observed frame) based on the co-added spectrum
  described in \S3 are listed below. Note that the last two fluxes are obtained using the
  extrapolation of the best fit \baselinereflection\ model (see below), and are only illustrative. }

\begin{itemize}
\item  $F_{\rm(2 - 10keV)} =(7.92\pm0.15)\times10^{-13}\ergpcmsqps$. 
\item  $F_{\rm(0.3 - 10keV)} =(1.45\pm0.06)\times10^{-12}\ergpcmsqps$.
\item  $F_{\rm(0.3 - 100keV)}=(3.37\pm0.23)\times10^{-12}\ergpcmsqps$. 
\item  $F_{\rm(10 - 40keV)} =(1.21\pm0.09)\times10^{-12}\ergpcmsqps$.
\end{itemize}

\noindent
Throughout the text, we utilize 2 models to describe the observed X-ray spectrum, which may be
described in \textsc{xspec} as follows:

\begin{itemize}
\item {\textit{Baseline-simple}: \textsc{phabs $\times$ (zphabs  $\times$ (zpowerlaw + diskbb + relline))}}
\item {\textit{Baseline-reflection}: \textsc{phabs     $\times$ (zphabs $\times$ (zpowerlaw + relconv $\otimes$ reflionx))}}, 
 \end{itemize} 
  \noindent where $\times$ and  $\otimes$ indicate multiplication and convolution respectively. 

Optical studies\cite{DaiKochanek2010quasar} have established the size of the optical disk in \rx\ to
the order of $100$\rg.  Microlensing studies in X-rays have subsequently constrained the
size\cite{DaiKochanek2010quasar, size1104, ChartasKochanek2012quasar} of the X-ray emitting region
to $\lesssim10$\rg.  As the accretion disk around a supermassive black hole emits mostly in
optical/UV, the X-ray emitting region constrained by these studies to be $\lesssim10$\rg\ is
traditionally associated with the corona. However, we show in this work that at least part of this
emission is due to reprocessed X-rays in the innermost regions around the black hole. Thus, the
$\sim10$\rg\ upper limit found in microlensing studies is likely to be characteristic of the size of
the reprocessing region\cite{reis2013corona}, with the corona actually limited to a region
$\ll10$\rg.

All calculations in this paper assume a flat $\Lambda{\rm CDM}$ cosmology with $H_{o}=70\kmpspmpc$,
$\Omega_{vac}=0.73$ and $\Omega_{M}=0.27$. All figures in this manuscript are shown in the observed
frame.\\

\newpage
\section{Toward Black Hole Spin: Chandra}
Due to the lensed nature of this source we are afforded a remarkable number of observations of the
quasar despite a modest number of pointings. Ref\cite{Tewes2012} estimated the time delay between
images B and C to be at most $\approx0.3$~days and the delay between B and D to be between
$\approx90-96$~days. As such, each of the 30 individual pointings effectively provides up to 4
spectra probing different epochs.  In order to account for possible intrinsic variability present in
the large number of spectra of \rx, we proceed by modelling the various spectra using similar
methodologies to those often employed in the analyses of X-ray binaries where large sets of pointed
observations are available (see e.g. ref\cite{reis20121650}).

As discussed thoroughly in\cite{ChartasKochanek2012quasar}, the variability of Images-B and C
through all the epochs are thought to be closely related to the intrinsic variability of the
quasar. Those authors estimate that the intrinsic variability of the quasar should be no larger than
28\%.  Images A and D on the other hand display a high level of variability which is attributed to
microlensing.  In the following, we fit all 27 epochs of Image-B with a physically motivated model.\\

\subsection{Fits with phenomenological models: \textit{Baseline-simple}}
\addcontentsline{toc}{subsection}{3.1.  Fits with phenomenological models: \textit{Baseline-simple} }
\subsubsection{Individual image-B spectra:}
\addcontentsline{toc}{subsubsection}{3.1.1. Individual Image-B Spectra}

We start by confirming the presence of the soft-excess and possible residuals around the iron line
region for all 27 spectra of image-B. We do this again in a phenomenological manner by using the
\textit{baseline-simple} model, allowing only the normalisations of the various components as well
as the power-law indices to vary between each epochs. We again constrain the rest frame energy of
the relativistic line to the 6.4-6.97\kev\ range and for simplicity assume that this is not varying
between epochs. Extended Data Fig.~1 (a) shows the best fit ($\chisq/\nu=2207.2/2172=1.02$) models
for the simultaneous fit of all 27 observations. On the right, we again remove the \diskbb\ and line
component to emphasise the presence of the soft-excess and the iron line.  We also show over-plotted
on the data in Extended Data Fig.~1 (b), the ratio between the total model and the illuminating
power-law for the spectrum of epoch-23. Replacing the relativistic line with a simple Gaussian
profile with energies constrained in a similar manner to \relline\ and again allowing the
normalisation to vary between epoch, resulted in a worse fit ($\Delta\chisq = +32.9$ for two fewer
degrees of freedom) compared to the \textit{baseline-simple}~model.

When using the \textit{baseline-simple} model here, we have made the logical assumption that the
inner disk inclination (measured with respect to the normal of the disk where 0 and 90 degrees mean
face-on and edge-on, respectively) and the black hole spin are not changing. Furthermore, at this
early stage in the analysis we have also tied the ionisation state of the disk as well as the disk
emissivity profile between the different epochs.  With these basic assumptions in mind, the
\textit{baseline-simple} model yields a spin and inclination of $a=0.86^{+0.06}_{-0.08}~{\rm
  Jc/GM^2}$ and $\theta=17^{+8}_{-3}$ degrees respectively (90\% confidence), as well as an average
emissivity $q>6.2$ for the image B data. Our results also indicate that there is no large neutral
column at the source redshift, with an upper limit based on this model of $N_{\rm H}
(z=0.658)<6.5\times 10^{20}\pcmsq$. Similar conclusions for the low intrinsic absorption and low
inclination were found by ref\cite{ChartasKochanek2012quasar}.  We also note that the mean value for
the powerlaw index is $\Gamma= 1.61\pm0.11$ (s.d.), and we find no \textit{statistically
  significant} variation in this parameter through all epochs. Similar conclusions regarding the
constancy of $\Gamma$ in Image-B were made by ref\cite{size1104}, where the authors find an average
value of $\sim1.68$ ranging from $\sim1.41- 1.82$ based on the first six epochs (see their Table~3).

We note briefly that despite the increase in the number of spectra from 3, as considered previously
when focusing on just epoch 23, to 27 for the full image B dataset, the total degrees of freedom
does not increase by a similar factor since the exposures and thus S/N of the various observations
are not the same. In fact, $\nu$ goes from $\sim 405$ to $\sim2172$, a factor of $\sim5$ increase.\\

\subsubsection{Combined ``microlensing-quiet'' images-B and C:}
\addcontentsline{toc}{subsubsection}{3.1.2. Combined ``microlensing-quiet'' images-B and C:}

It is common practice in the study of nearby Seyferts to use a single time averaged-spectrum when
performing detailed analyses aimed at obtaining the spin parameter; similar to the goal set
here. This practice is motivated in large part by either the computational intensity of such tasks
or due to low S/N in individual spectra. However, the time averaged result has usually been shown to
be consistent with that obtained through more detailed analyses, e.g. time resolved spectroscopy,
when it has been possible to assess both. A case in point is the measured spin parameter of NGC~3783
where work has been done on both time resolved and averaged spectra to arrive at similar value for
the spin\cite{3783p1, reis3783}, using a methodology similar to that employed here.

As detailed in ref\cite{ChartasKochanek2012quasar}, Images-A and D are thought to be representative
of microlensing ``active" states, meaning that the observed variability is largely due to
microlensing effects. As microlensing is capable of selectively amplifying different regions
depending on their size, it is possible that spectra in the active states are deformed in a complex
manner\cite{popovic01, Abajas02} unrelated to General relativistic and reprocessing effects from the
inner accretion disk\cite{rossfabian1993}. As such, in this section we begin by co-adding the
spectra and responses of Images-B and C to form two time-averaged spectra representative of the
microlensing ``quiet" state. We fit these spectra in the 0.4-8.0\kev\ range as a conservative
precaution, as the lowest energy bins are systematically above any reasonable continuum fit, likely
related to known ACIS calibration issues, e.g., see
\\ \href{http://cxc.harvard.edu/ciao4.4/why/acisqecontam.html}{http://cxc.harvard.edu/ciao4.4/why/acisqecontam.html}
and\\ \href{http://cxc.harvard.edu/cal/Acis/Cal_prods/qeDeg/index.html}{http://cxc.harvard.edu/cal/Acis/Cal\_prods/qeDeg/index.html}.

Extended Data Fig.~2 (a) shows the co-added spectra of Images-B and C.  It is clear from the
residuals as well as the poor $\chisq_{\rm B}/\nu_{\rm B}=559.6/358=1.56$ and $\chisq_{\rm
  C}/\nu_{\rm C}=291.6/250=1.17$ for B and C respectively, that a simple power-law is not a good
representation of the spectra. Adding a \diskbb\ component for the soft excess again improves both
fits ($\chisq_{\rm B}/\nu_{\rm B}=431.0/356=1.21$; $\chisq_{\rm C}/\nu_{\rm C}=262.9/248=1.06$) but
clear residuals remains above $\sim2$\kev.
component again constrained to lie between 6.4 and 6.97\kev\ and initially having a powerlaw
emissivity profile. This improved the fit ($\Delta\chisq_{\rm B}/\Delta\nu_{\rm B}=-27.6/-5$;
$\Delta\chisq_{\rm C}/\Delta\nu_{\rm C}=-13.7/-5$).  However, there still appeared to be narrow
residuals at 3.86\kev\ (6.4\kev\ in the rest frame) in both spectra (see bottom panel of Extended
Data Fig.~2). This possible emission line could be associated with reflection from distant material
as is often found in nearby Seyferts, or it could be coming from the outer parts of the accretion
disk. Initially assuming the latter, we change the emissivity profile of \relline\ so that within a
radius of 10\rg\ the emissivity is $Q_{in}$ and beyond this radius it is described as $Q_{out} > 2
$.  Such a broken powerlaw prescription for the emissivity profile is naturally expected when one
suspects the black hole to be rapidly rotating and the corona to be
compact\cite{FabZog09,Fabian2012cyg,wilkins2011}. The break at 10\rg\ is motivated both
theoretically\cite{Fabian2012cyg} as well as observationally since this is the likely scale for the
X-ray emitting region in \rx\ as measured using gravitational
microlensing\cite{size1104,DaiKochanek2010quasar,ChartasKochanek2012quasar}; however, we note that
allowing this parameter to be free does not change the results presented here as the break radius is
not very well constrained.

This \textit{baseline-simple} model with a broken powerlaw emissivity profile provides a good fit to
the time-averaged spectra of both images ($\chisq_{\rm B}/\nu_{\rm B}=393.9/350=1.13$; $\chisq_{\rm
  C}/\nu_{\rm C}=231.0/242=0.95$). With the increased S/N provided by the co-added spectrum, the
addition of the relativistic line component is now significant at $>4.3\sigma$ level of confidence
(F-test false alarm probability of $1.9\times 10^{-5}$) for image-B, and at the $4\sigma$ level
(F-test false alarm probability of $7.4\times 10^{-5}$) for image-C. Extended Data Table~1
summarises the various parameters. Most importantly, the value for the spin found for both images
(see Extended Data Fig.~3; black curves) are in excellent agreement with the results found in the
previous section for the time-resolved fits of Image-B. The inclination and the power-law index
found for image-B are also in excellent agreement with the results for the time-resolved analyses
presented in the previous section.

As mentioned previously, a further possibility for the narrow component seen in Extended Data Fig.~2
is emission from distant material. Indeed it is possible that the narrow feature is due to a
combination of these two effects, i.e., emission from distant material and a broken emissivity
profile. Such a fit combining both a broken emissivity profile with $Q_{out} = 3$ -- the expected
asymptotic value at large ($\gtrsim10$\rg) distances -- together with a Gaussian to characterise
distant reflection having a width frozen at 1\ev, also provides a satisfactory fit with $\chisq_{\rm
  B}/\nu_{\rm B}=400.2/350=1.14$ and $\chisq_{\rm C}/\nu_{\rm C}=231.8/242=0.96$. The confidence
contour for this model are also shown in Extended Data Fig.~3 (blue curves). All parameters remain
essentially unchanged from those presented in Extended Data Table~1.

Before applying the more physically-motivated \textit{baseline-reflection} model to these spectra,
it is worth comparing our results for Image-C with those obtained by
ref\cite{ChartasKochanek2012quasar}. In their work, they model the co-added spectrum of Image-C with
a single power-law ($\Gamma=1.79\pm0.02$) together with a Gaussian at $E_{\rm
  Ga}=6.36^{+0.07}_{-0.08}$\kev\ (rest frame) having an equivalent width of
$EW=154^{+70}_{-80}$\ev. We find that a similar model indeed provides an adequate fit with
$\chisq_{\rm C}/\nu_{\rm C}=260.3/247=1.05$, $E_{\rm Ga}=6.38^{+0.05}_{-0.06}$, $EW=137$\ev, and
$\Gamma=1.76\pm0.03$, meaning that all our results for this model are in excellent agreement with
their work. Although this model does formally provide an adequate fit to the time-averaged data of
Image-C, the addition of a soft component again improves the fit dramatically ($\Delta\chisq_{\rm
  C}/\Delta\nu_{\rm C}=-23.2/-2$) and an F-test shows that this extra component is required at the
$4.4\sigma$ level (F-test false alarm probability of $1.1\times 10^{-5}$).  Although we cannot
differentiate between the \relline\ + \diskbb\ model ($\chisq_{\rm RelLine}/\nu_{\rm
  relLine}=231.0/242=0.95$) from the Gaussian + \diskbb\ model ($\chisq_{\rm Gaussian}/\nu_{\rm
  Gaussian}=237.2/245=0.97$) on a statistical basis for image-C alone, the presence of the line
\textit{together with} the extra soft component is statistically robust and for the various reasons
presented throughout this manuscript we favour the relativistic reflection based interpretation for
the emission line seen in Image-C.\\

\subsubsection{Combined ``microlensing-active'' images-A and D:}
\addcontentsline{toc}{subsubsection}{3.1.3. Combined ``microlensing-active'' images-A and D:}

We now consider the two ``microlensing-active'' images, A and D. Ref\cite{ChartasKochanek2012quasar}
highlight two periods of distinct behaviour in the evolution of Image-D, one in which the the
Fe-\ka\ line profile appears to show a distinctive peak at $\sim6.4$\kev, which they call Periods~1
and 3 (epochs 1-16 and 23-30 respectively), and the other where the Fe-line region is better
described by a double profile during their Period~2 (epochs 17-22). In comparison, they did not see
any such evolution from the microlensing-quiet period in Image-C. Extended Data Fig.~2 (b) shows the
residuals to a model consisting of a combination of power-law and Gaussian profiles in a similar
manner to that of ref\cite{ChartasKochanek2012quasar}. We cannot directly compare our residuals to
those of ref\cite{ChartasKochanek2012quasar} since the authors did not show such a figure; however,
close inspection of their spectra indicates residuals similar to the ones shown here.

It is clear from the residuals below $\sim1$\kev\ shown in Extended Data Fig.~2 (b), that this
simple power-law does not provide a good representation of Periods 1 and 3 for either Image-A or
D. However, possibly owing to the poor S/N afforded in Period-2, the data is here consistent with a
simple power-law and as such we do not consider these data further. The familiar appearance of the
soft excess is again well characterised phenomenologically by a \diskbb, and the addition of such a
component to the co-added spectrum of Image-D, Period 1+3 yields an improvement of
$\Delta\chisq_{\rm D13}/\Delta\nu_{\rm D13}=-31.9/-2$ (final $\chisq_{\rm D13}/\nu_{\rm
  D13}=238.5/242=0.99$).  Similarly, adding a \diskbb\ to the corresponding period of Image-A yields
$\chisq_{\rm A13}/\nu_{\rm A13}=250.4/267=0.94$, an improvement in $\chisq$ of 58.1 for 2 degrees of
freedom.  The residuals during Periods-1 and 3 for both Images-A and D closely resemble those of the
microlensing-quiet Images B and C (see Extended Data Fig.~2) and the model we have used to account
for the soft excess can again be linked with the clear presence of Fe emission.\\

\subsubsection{Summary of phenomenological (\textit{Baseline-simple}) Results:}
\addcontentsline{toc}{subsubsection}{3.1.4. Summary of Phenomenological (\textit{Baseline-simple}) Results:}

We have used the \baselinesimple\ model to provide initial constraints on the spectral shape of \rx,
and to compare the results obtained from the different individual images. First of all, we stress
that the soft excess and Fe K residuals present in epoch-23 and detailed above are present in all 27
epochs of Image-B (Extended Data Fig.~1). We find no significant evolution in the powerlaw index as
obtained with the \baselinesimple\ model throughout these epochs. The average value is found to be
$\Gamma = 1.61\pm0.11$~(s.d.). A joint fit to all 27 epochs suggests the lack of any significant
absorption at the source redshift, and gives a spin parameter $a=0.86^{+0.06}_{-0.08}$~ (90\%
significance). Co-adding the spectra of image-B, the soft-excess (modeled with a \diskbb) and the
broad iron line are required at $>>5\sigma$ and $>4.3\sigma$ respectively, even before the data from
images A, C and D are considered.

The co-added spectra of images A, C and D (after excluding the powerlaw dominated Period 2 for
images A and D) show similar residuals to a simple powerlaw continuum as the co-added spectrum of
Image-B (Extended Data Fig.~2). The spin parameter obtained with the \baselinesimple\ model for
image C is fully consistent with that obtained previously for Image-B (See Extended Data Table~1 and
Extended Data Fig.~3). Unfortunately, owing to the lower S/N in the co-added spectra from images A
and D, these data do not allow for individual spin constraints, but we again stress that similar Fe
K residuals to those seen in images B and C are observed.\\

\subsection{Fits with physical models: \textit{Baseline-reflection}:}
\addcontentsline{toc}{subsection}{3.2.  Fits with physical models: \textit{Baseline-reflection} }
We now proceed to fit the spectra with a self consistent physical model.  As mentioned in the main
manuscript, the use of a disk blackbody for the soft excess is purely phenomenological, and is only
used to model the soft-excess in a similar manner to previous work on quasars for ease of direct
comparison (see e.g. \cite{Porquet04quasars,Piconcelli05quasars}).

We replace the \diskbb\ and \relline\ components with the reflection model \reflionx\ of
\cite{reflionx} and account for relativistic affects using the \relconv\ kernel from\cite{relconv}.
To the best of our knowledge, \relconv\ (and its equivalent line model \relline) {represent the
  current state of the art in relativistic reflection modelling.}  In \xspec\ terminology, this
combination of components
reads:\\ $\zphabs\times\phabs\times(\zpo+\relconv\otimes\reflionx)$\\ where $\otimes$ denotes
convolution.
 
When using \reflionx, we constrained the power-law index of the reflection component to be that of
the illuminating power-law and set its redshift to that of the quasar. The iron abundance is
initially frozen at Solar (Fe/solar=1).\\

\subsubsection{Individual image-B spectra:}
\addcontentsline{toc}{subsubsection}{3.2.1. Single images}

We begin our self-consistent analysis by modeling the 27 spectra from image-B simultaneously. The
intrinsic neutral absorbing column, the black hole spin, the accretion disk inclination and the
emissivity index are linked between all 27 spectra, i.e. assumed to be constant with time, while the
normalisations of \reflionx\ and the power-law, the photon index and the disk ionisation were
allowed to vary between them.

The \baselinereflection\ model characterised all the data well ($\chisq/\nu=2241.8.6/2174=1.03$),
including the intrinsic variability, in a self-consistent manner.  \rx\ is thought to be
accreting\cite{Sluse2012mass} at $L/L_{\rm Edd}\sim0.07$, -- where $L_{\rm Edd}$ is the Eddington
limit -- which is similar to the accretion rate often observed in a number of Seyferts, including
the canonical source for reflection based spin measurements:
MCG-6-30-15\cite{tanaka1995,Fabianvaughan03,Fabian02MCG,kerrconv}. Therefore, it is no surprise that
this combination of model works so well for \rx.

We show in Extended Data Figure~3 the confidence limits for the spin as obtained from the combined
statistics of these 27~observations for a total exposure of $\sim318$\ks. The spin is constrained to
$a=0.90^{+0.08}_{-0.10}~{\rm Jc/GM^2}$ at the $3\sigma$ level of confidence (99.73\%). The
inclination, emissivity index and intrinsic absorption are constrained to $<19$~degrees,
$Q=5.2\pm0.3$ and $N_{\rm H} = (1.0^{+0.2}_{-0.1})\times 10^{21}\pcmsq$ at the 90\% level.

We stress here that the constraint on the spin does not come solely from the iron emission feature,
but from the full reflection spectrum, including the featureless soft-excess. Of course another way
to obtain a featureless continuum is to have metallicities significantly below the solar value used
here. This is highly unlikely as quasars are famously known to have enhanced metallicities
\cite{HamannFerland1993}, with a near flat evolution from $z=0$ up to $z\sim4-5$, after which it is
possible that it declines to solar or even subsolar values\cite{Iwamuro2002}.\\

\subsubsection{Combined ``microlensing-quiet'' images-B and C:}
\addcontentsline{toc}{subsubsection}{3.2.2. Combined ``microlensing-quiet'' images-B and C:}

Following our phenomenological analysis, we now again consider the data from the two
``microlensing-quiet" images, B and C. Here, though, we limit our analysis to the co-added spectra
obtained from these images, as simultaneously modeling the individual observations of both images
would be extremely computationally intensive. As discussed previously, these co-added spectra probe
the time averaged features of the system, in an identical manner to that often exploited in nearby
Seyferts.

Again we apply the \baselinereflection\ model to the two combined spectra.  A narrow (1\ev) Gaussian
is included at 6.4\kev\ and we use a broken emissivity profile with $Q_{out} > 2$.  Extended Data
Table~1 details the parameters found for this model. The confidence contours for the spin as
obtained for each individual Image are also shown in Extended Data Figure~3 (red curves). It is
clear from Extended Data Table~1 and Fig.~3 that the parameters between both Images-B and C are all
consistent with one another. The consistent shape of Image-C during the full observation was also
highlighted by ref\cite{ChartasKochanek2012quasar}. In addition, the spin constraints obtained from
each image are consistent with that found previously during our time resolved analysis of image B.\\

\subsubsection{Combined ``microlensing-active'' images-A and D; Periods 1 and 3:}
\addcontentsline{toc}{subsubsection}{3.2.3.  Combined ``microlensing-active'' images-A and D; Periods 1 and 3:}
 
Finally, we consider the ``microlensing-active'' images (A and D) in the context of the
\baselinereflection\ model. The \textit{exact same} model for the co-added spectrum of Image-B
detailed in Extended Data Table~1 gives $\chisq_{\rm A13}/\nu_{\rm A13}=253.8/274=0.93$ when applied
to the co-added spectrum from image A and simply renormalised. This is as expected since Images-A
and B (and C) are probing similar times. However, as Image-D lags the rest by $\sim100$~days, we use
the same model as above but allow the various normalisations, inner disk emissivity profile,
power-law index and disk ionisation to vary. This again provides an excellent fit to the co-added
spectrum of Image-D ($\chisq_{\rm D13}/\nu_{\rm D13}=245.56/243=1.01$). All parameters are
consistent within errors with those reported in Extended Data Table~1, although we note that the
power-law index and the disk ionisation are not particularly well constrained ($\Gamma_{D13}<2.2$;
$\xi_{D13}= 1270^{+1100}_{-950}\ergcmps$).\\

\subsubsection{Summary of self-consistent (\textit{Baseline-reflection}) results:}
\addcontentsline{toc}{subsubsection}{3.2.4. Summary of self-consistent (\textit{Baseline-reflection}) results:}

In addition to our phenomenological analysis, we have also considered the data obtained from the
individual \chandra\ images in the context of the physically self-consistent
\baselinereflection\ model. We began by applying this model to the 27 individual spectra of image B
simultaneously, and then to the co-added spectra obtained from images B and C. Excellent fits are
obtained in each case. The spin constraints (Extended Data Fig.~3) obtained from these analyses are
-- at the 90\% confidence level -- $a=0.90^{+0.04}_{-0.05} ~{\rm Jc/GM^2}$ (image B, time resolved),
$0.92^{+0.04}_{-0.06}$ (image B, co-added) and $0.80^{+0.08}_{-0.06}$ (image C, co-added), which are
all consistent with one another. Finally, the \baselinereflection\ model also provides excellent
fits to the co-added spectra from images A and D, and consistent results are again obtained.\\

\newpage
\section{The Spin of \rx} 
\subsection{Combined Chandra data: 1.13~Msec of exposure:}
\addcontentsline{toc}{subsection}{4.1. Combined Chandra data: 1.13~Msec of exposure}
We have shown above that the spectral shape of the microlensing quiet images (Images B and C) as
well as that for the more active images during certain periods (Images-A and D during periods 1 and
3 of ref\cite{ChartasKochanek2012quasar}) are extremely similar to one another. Following the
standard procedure employed in the study of nearby Seyferts, we now combine these observations into
a single time-averaged spectrum.  This combined spectrum has a total exposure of 1.13~Msec and $\sim
100,000$~counts in the 0.4-8.0\kev\ range.\\

\subsubsection{Baseline-simple:}
\addcontentsline{toc}{subsubsection}{4.1.1. \textit{Baseline-simple}} Extended Data Fig.~4 (a) shows
the time-averaged \chandra\ data fit with the\textit{ baseline-simple} model. This results in an
excellent fit with $\chisq/\nu=410.4/408=1.006$. The relativistic line at $E_{\rm RelLine} =
6.49^{+0.05}_{-0.04}$\kev\ returns a broken emissivity profile with $Q_{\rm in} = 7.2^{+0.4}_{-0.9}$
and $Q_{\rm out} < 2.3$. Recall that broken emissivity profiles similar to the one found here is
naturally expected when one suspects the black hole to be rapidly rotating and the corona to be
compact\cite{FabZog09,Fabian2012cyg,wilkins2011}.  The power-law index is found to be
$1.69^{+0.02}_{-0.04}$ and the spin is constrained to $$a=0.82^{+0.05}_{-0.09}~ {\rm (90\% ~
  confidence).}$$ Figure~2 in the main manuscript shows the ratio to this model after setting the
normalisation of the disk and relativistic line component to zero, in order to highlight the
contribution from these components. We also show in Extended~Data~Fig.~4 (b), the residuals to a
single power-law before (top; $\chisq/\nu=782.0/416=1.88)$ and after the addition of a
\diskbb\ component (bottom; $\chisq/\nu=546.3/414=1.32$). An equally good fit can be achieved with
the addition of a narrow (1~eV) Gaussian at 6.4\kev\ together with the relativistic line now having
$Q_{\rm out} =3$ ($\chisq/\nu=411.9/408=1.01$), and the spin value remains unchanged.\\

\subsubsection{Baseline-reflection:}
\addcontentsline{toc}{subsubsection}{4.1.2. \textit{Baseline-reflection}} As a final step in our
assessment of the robustness of our results based on the \chandra\ data alone, we again replace the
phenomenological combination of components described above with \reflionx\ convolved with
\relconv. A narrow (1ev) Gaussian is again added at 6.4\kev\ and we use the broken emissivity
profile with $Q_{out} = 3$, as described above. Extended~Data~Fig.~5 (a) shows the fit to the
time-averaged data and the right panel shows the extrapolated model. This model self-consistently
accounts for the broad iron line and the soft-excess seen in this $z=0.658$ quasar, in a manner
similar to the canonical recipe for nearby sources. The best fit ($\chisq/\nu=439.6/409=1.07$) again
yields constraints on the spin which are consistent with all others presented in this work,
i.e.  $$a=0.90^{+0.07}_{-0.15}~({\rm 3\sigma ~ confidence).}$$We show the confidence contour for
this model in Extended~Data~Fig.~3 (panel c; magenta contour). Extended Data Table~2 shows the
parameters for this final model. As opposed to the phenomenological model described above, the
additional narrow Gaussian at 6.4\kev\ is moderately statistically significant ($\Delta\chisq=-6.5$
for one more degree of freedom) when using the physically motivated model. As such, it appears that
the narrow feature at 6.4\kev\ is indeed due to a combination of emission from distant materials as
well as from the outer regions of the accretion disk.\\

\subsection{Additional XMM-Newton data:}
\addcontentsline{toc}{subsection}{4.2. Additional XMM-Newton data}
\subsubsection{Baseline-simple:}
\addcontentsline{toc}{subsubsection}{4.2.1. \textit{Baseline-simple}}
We now also consider our recent observation of RX\,J1131-1231 with \xmm. A simple absorbed powerlaw
model does not adequately describe the 0.3-10\kev\ range of the \xmm\ \epicpn\ spectrum, with
$\chisq/\nu=982.5/858=1.15$ (see Extended~Data Fig.~6). Adding a disk component similar to the
\baselinesimple\ model improves the fit significantly ($\Delta\chisq/\Delta\nu=65/2$) and an F-test
shows that this extra component is required at greater than the $7\sigma$ level of confidence
(F-test false alarm probability of $2\times 10^{-13}$). The relatively high effective area of
\xmm\ at energies greater than 5\kev\ also allows for the clear detection of a break in the
continuum. Indeed by allowing the powerlaw to break, we find $\chisq/\nu=899.9/854=1.05$, an
improvement of $\Delta\chisq/\Delta\nu=17.7/2$ (F-test false alarm probability of $2.5\times
10^{-4}$) over the fit without a break. Finally, the addition of a narrow Gaussian at
$6.48^{+0.04}_{-0.03}$\kev\ (rest frame) yields a best fit of $\chisq/\nu=862.4/852=1.01$ (
$\Delta\chisq/\Delta\nu=37.5/2$; F-test false alarm probability of $1.3\times 10^{-8}$) for this
phenomenological combination of components. This final model suggests the presence of a soft excess
which can again be characterised by a \diskbb\ component with $kT=0.22\pm0.03\kev$, a powerlaw with
an index $\Gamma=1.83^{+0.07}_{-0.03}$ up to a break at $E_{\rm break}=5.5^{+0.5}_{-2.2}\kev$, at
which point the continuum hardens to $\Gamma=1.28^{+0.33}_{-0.19}$.  We show in Fig.~2 of the main
manuscript the ratio to this $\Gamma=1.83$ powerlaw.

Within the reflection-paradigm, this combination of a hardening at $\sim10\kev$ (rest frame)
together with a soft excess below $\sim2\kev$ can be characterised as the beginning of the Compton
hump and the blending of soft emission lines, respectively. To our knowledge, this is the first
clear detection of a break associated with a Compton hump in a moderate-\textit{z} quasar. In the
following subsection, we proceed by modelling this spectrum within the context of relativistic
reflection.\\

\subsubsection{Baseline-reflection:}
\addcontentsline{toc}{subsubsection}{4.2.2. \textit{Baseline-reflection}}

We replace the phenomenological \diskbb\ component as well as the broken powerlaw with the
\baselinereflection\ model together with a narrow Gaussian (see Extended~Data~Fig.~6; b). This model
is detailed in Extended Data Table~2. With a best fit of $\chisq/\nu=849.4/849=1.000$, it is clear
that this self-consistent description provides a quality of fit that significantly outperforms even
that of the phenomenological combination presented above.

We show in Extended~Data~Fig.~3 (panel c; green) the confidence range for the spin as obtained from
the \xmm\ data alone. The spin found here of $$a=0.64^{+0.33}_{-0.14}~({\rm 3\sigma ~
  confidence),}$$ is again consistent with the range found during our analysis of the
\chandra\ data.

It is clear from Extended Data Table~2 and Extended~Data~Figs.~5 and 6, that the ratio of the
reflected flux to the powerlaw flux (the reflection fraction) is lower during the \xmm\ observation,
but the overall flux is higher. This trend of decreasing reflection fraction with increasing flux is
often observed\cite{Vaug04} in local AGN and within the reflection/light-bending
interpretation\cite{Miniu04} involves a corona whose height above the accretion disk is changing. In
this scenario, a corona that is relatively close (a few \rg s) to the black hole will have more of
its emission bent towards the disk, decreasing the fraction of the coronal emission that escapes to
the observer, and thus increasing the observed reflection fraction. On the other hand, if the corona
is further from the black hole so that its emission can be better characterised as being isotropic,
then the total flux illuminating the disk decreases (more flux escapes to the observer) and so does
the reflection fraction. Importantly, the behaviour seen here is not only fully consistent with the
expectations of gravitational light-bending, but it also \textit{requires} a system having a high
spin, consistent with the value reported in this work.  Similar behaviour has been reported for a
number of AGN, most famously 1H0707-495\cite{FabZog09} and MCG -6-30-15 \cite{Vaug04} as well as
stellar mass black holes\cite{reis20121650}.\\

\subsection{Joint XMM-Newton and Chandra fit with \textit{Baseline-reflection} model:}
\addcontentsline{toc}{subsection}{4.3.   Joint XMM-Newton and Chandra fit with \textit{Baseline-reflection} model}

It is clear from Extended Data Table~2 that, where the parameters are not expected to vary between
observations, i.e. inclination, column density, and spin, the \xmm\ observation yields consistent
parameters to those obtained from the time-averaged \chandra\ data. In order to optimise the S/N and
obtain a final estimate of the spin parameter of \rx, we proceed by fitting both data sets
simultaneously with the \baselinereflection\ model, with the inclination, column density and spin
tied between them.  Extended Data Table~2 also details the various parameters for this final, joint
fit and Figure~3 in the main manuscript (duplicated in Extended~Data~Fig.~3; panel c; black) shows
the confidence contour obtained for the spin where we find a value of $$a=0.87^{+0.08}_{-0.15}~({\rm
  3\sigma ~ confidence),}$$ based on the combined \chandra\ and \xmm\ data.\\

\newpage
\section{No Evidence for Complex Absorption in \rx:} To this point, we have
based our modelling on the reflection paradigm and followed the well established methodology that
has been applied to many local Seyferts. We note, however, that $\sim 50\%$ of these
Seyferts\cite{reynolds97wa} and a similar number of quasars\cite{Porquet04quasars} also display
evidence for absorption by partially ionised, optically-thin material local to the accretion flow
(``warm'' absorbers; WAs).

Indeed, the canonical Seyfert galaxy, MCG-6-30-15, displays one of the most prominent relativistic
lines known, and its X-ray spectrum also requires the presence of multiple absorption zones. Early
spin measurements of MCG-6-30-15 (e.g. ref\cite{tanaka1995}) often strictly focused on data
$\gtrsim2$\kev, as the main effect of these warm absorbers are below this energy (e.g.
ref\cite{Lee2001ApJ}), notably the two strong edges of {\rm O~VII and O~VIII} at $\sim
0.74$\kev\ and $\sim0.87$\kev, respectively \cite{Fabian1994PASJ} (note that for \rx\ at $z=0.658$,
this restricts the bulk effect of any possible WA to energies $\lesssim0.5$\kev). However,
subsequent detailed analyses accounting for the multizone warm absorber present in this source still
obtained consistent spin measurements\cite{Vaug04, kerrconv, Chiang2011} and concluded that the
relativistic iron line is robust to the precise details of the WA
(e.g. \cite{Younglee2005}). Nonetheless, the presence of such a component in half of all quasars
prompted us to investigate whether the residuals seen in \rx\ could be explained by WAs, and what
effect the presence of a putative WA will have on our ability to constrain the spin of the black
hole.\\

\subsection{Partially Ionised Absorption:}
\addcontentsline{toc}{subsection}{5.1. Partially Ionised Absorption}

\subsubsection{Phenomenological modeling:}
\addcontentsline{toc}{subsubsection}{5.1.1. Phenomenological modeling}
We start by fitting the co-added spectra of Images-B and C with a model describing partially
covering absorption by a partially ionised medium (\zxipcf\ in \xspec; ref\cite{zxipcf}). The WA is
characterised with an ionisation parameter $\xi_{\rm wa}=L/nr^2 \ergcmps$, where $L$ is the ionising
X-ray luminosity ($\ergps$), $n$ the gas density (${\rm cm^{-3}}$), and $r$ is the distance in
centimeters between the source of ionising X-rays and the absorbing gas. The model also includes the
covering fraction ($cf$) which defines the fraction of the source which is covered by the absorbing
gas with a column density $N_{\rm H;wa}$, while the remaining (1-$cf$) flux from the source escapes
directly to the observer. We initially allow the ionisation parameter, column density and covering
fraction of \zxipcf\ to be free, and apply this to a simple absorbed power-law model. In
\xspec\ terminology this model reads \textsc{phabs$\times$ (zphabs $\times$(zxipcf $\times$
  (zpowerlaw)))}.

This model provided a goodness of fit equal to that of the power-law+\diskbb\ combination for both
images ($\chisq_{\rm B}/\nu_{\rm B} = 426.1/355=1.20$ and $\chisq_{\rm C}/\nu_{\rm C}
=263.6/247=1.07$ for Images-B and C respectively), meaning that it can account for the
``soft-excess" to the same degree as the previous model using \diskbb. However, like the fit with a
\diskbb, this model still cannot account for the residuals above $\sim2$\kev. Adding a Gaussian,
constrained to lie in the Fe K-shell energy range (6.4-6.97\kev\ local frame), to Image-B does not
improve the fit; however, upon lifting this constraint we obtain an improvement ($\chisq_{\rm
  B}/\nu_{\rm B} = 414.0/352=1.18$).  The Gaussian line has a centroid energy of $E_{\rm
  Gaussian}=3.61^{+0.26}_{-0.24}$\kev\ (local frame) and a width of $\sigma =
710^{+370}_{-570}$\ev. It is clear that such a broad line, with a centroid energy much lower than
the 6.4\kev\ expected for neutral iron is a rather unphysical combination which is artificially
mimicking a broad, relativistic line.  Alternative scenarios such as Compton broadening aimed at
explaining broad lines of this magnitude have been shown to not be a viable alternative in Seyfert
galaxies\cite{Waltonreis2012}.  As such, we replace the Gaussian with \relline\ and constrain the
energy to 6.4-6.97\kev\ as per usual. This model, with a WA, provides a better fit ($\chisq_{\rm
  B}/\nu_{\rm B} = 407.2/349=1.17$), and most importantly, models the residuals in a physically
motivated manner. We note that adding a second zone does not improve the fits.  Focusing on Image-C,
we find that while the addition of a Gaussian does improve the fit ($\chisq_{\rm C}/\nu_{\rm C} =
241.0/244=0.99$), it is not statistically significant. Replacing this Gaussian with \relline\ does
remove clear systematic residuals and indeed increases the goodness of fit to $\chisq_{\rm
  C}/\nu_{\rm C} = 234.1/241$ ($\chi^2_{\nu} = 0.97$).

The ionisation parameter of the putative WA is not well constrained for either image, with both
cases resulting in upper limits of ${\rm log_{10} } \xi \lesssim 1.5$. It is clear that the WA is
having the same affect as the phenomenological \diskbb\ model. However, the spin obtained via the
relativistic line alone for Images-B and C are consistent with all other results presented here
($a_{\rm B}=0.89^{+0.02}_{-0.09}$ and $a_{\rm C}=0.8^{+0.09}_{-0.06}$ at the 90\% confidence), and
most importantly is still constrained to be high.\\

\subsubsection{Baseline-reflection:}
\addcontentsline{toc}{subsubsection}{5.1.2. \baselinereflection}

As a broad relativistic Fe-\ka\ line is naturally accompanied by other emission at lower energies,
which can self-consistently account for the soft-excess, we proceed by reverting back to our
\textit{baseline-reflection} model (as detailed on Extended Data Table~1) in order to model the
residuals above $\sim2$\kev (observed frame) in both images. However, we now include an additional
WA in order to investigate any possible effect on the results obtained with this model.

The addition of \zxipcf\ to the \baselinereflection\ model for the co-added Image-B data improves
the quality of the fit by $\Delta\chisq_{\rm B}/\Delta\nu_{\rm B}=-5.2/-3$ (final $\chisq_{\rm
  B}/\nu_{\rm B}=402.4/348=1.16$), i.e., this extra component is not statistically
significant. Nonetheless, we note that this fit to Image-B yields $N_{\rm H;wa;B}=
(1.4^{+5.0}_{-0.8} ) \times 10^{22} \pcmsq$, ${\rm log_{10} } \xi_{\rm wa; B} < 2.1$ and $cf=
0.2^{+0.3}_{-0.1}$. All other parameters stay the same as those presented in Table~S1, within the
errors.  As the addition of this extra component over the \textit{baseline-reflection} model is not
statistically significant, we also tried fixing ${\rm log_{10} } \xi_{\rm wa; B}= 2$, typical of
both Seyferts like MCG-6-30-15 as well as Quasars such as PG~1309+355
(ref\cite{Ashton2004wa}). Again the improvement over a model without such absorption is barely
significant at $\Delta\chisq_{\rm B}/\Delta\nu_{\rm B}=-3.1/-2$; nonetheless, this fit again gives a
low covering fraction ($cf<0.27$) and the constraint on the spin remains essentially unchanged
($a=0.91\pm0.05$) from that reported on Extended Data Tables~1. Freezing ${\rm log_{10} } \xi_{\rm
  wa; B} =3$ or 1 does not change this conclusion nor does the addition of a second WA.

Adding a similar absorber to the \textit{baseline-reflection} model for Image-C does not provide any
improvement over that reported on Extended Data Table~1 and none of the WA parameters are
constrained. We thus proceed by again freezing the ionisation to ${\rm log_{10} } \xi_{\rm wa; B}=
2$ and setting the column density to $N_{\rm H;wa;C}= 1 \times 10^{22} \pcmsq$ so as to obtain a
rough limit on any potential covering fraction. This imposed WA provides a fit that is statistically
similar to that shown in Extended Data Table~1 and sets a limit on the covering fraction of
$cf\leq0.12$. Again the spin remains unchanged.

We have also investigated the addition of a WA to the \textit{baseline-reflection} model used in the
time-resolved spectral analyses presented in \S~3. We initially allowed the column density and
covering fraction to vary between epochs but kept the ionisation parameter frozen at ${\rm log_{10}
} \xi_{\rm wa}= 2$. This did not improve the fits. Neither did freezing the ionisation at ${\rm
  log_{10} } \xi_{\rm wa}= 3$, or allowing it to vary between epochs. In fact, the additional 54 or
81 free parameters, depending on whether ${\rm log_{10} } \xi_{\rm wa}$ was frozen or not, made the
overall reduced $\chisq$ worse than that without this extra component. It is clear that the addition
of a WA does not improve the quality of the fits nor does it affect the values for the spin obtained
here. We therefore conclude that \rx\ does not require the presence of a WA on top of the
self-consistent \textit{baseline-reflection} model.\\

\subsection{Neutral Absorption:}
\addcontentsline{toc}{subsection}{5.2. Neutral Absorption}
We have also investigated the use of multiple neutral, partially covering absorption components, in
addition to the fully covering neutral absorption included in all models, with the \zpcfabs\ model
within \xspec. Initially we restrict our consideration to Compton-thin absorption. First, we add an
additional partially covering neutral absorber to the relativistic reflection model considered
previously for the joint \chandra\ and \xmm\ dataset. We obtain a best fit commensurate with that
reported on Extended Data Table~2, with $\chisq/\nu = 1279.2/1256=1.02$ and most importantly, a spin
parameter tightly constrained to $a=0.75^{+0.06}_{-0.09}$ ($3\sigma$); statistically similar to that
reported on Extended Data Table~2. Second, we attempt to construct absorption-dominated models, but
find that it is not possible to reproduce either the combined \chandra\ or the \xmm\ data solely
using reasonable combinations of such models, i.e. without any contribution from relativistic
reflection. For the \chandra\ data, residuals associated with the broad iron line still persist, and
it is not possible to account for the spectral curvature above $\sim$5\,\kev\ in the \xmm\ data.

Finally, we relax the requirement that the absorbers are Compton-thin. However, we find this only
influences the fit to the \xmm\ data, where we find that a partially-covering ($cf =
0.27^{+0.12}_{-0.11}$), Compton-thick absorber (\nh\ = $1.68^{+1.75}_{-0.72} \times 10^{24}\pcmsq$)
can successfully account for the high-energy excess in the \xmm\ data. We therefore also include in
this model a cold ($\xi=1$), unblurred \reflionx\ component aimed at simulating reprocessed emission
from Compton-thick material far from the black hole. Nevertheless, when this model is applied to the
joint \xmm\ and \chandra\ dataset, with the partially covering absorber allowed to vary between the
\xmm\ and \chandra\ data, despite a visibly good fit to the \xmm\ data above $\sim 5$\kev\ now being
obtained this model still provides a worse fit ($\Delta\chisq = 54.7$) than the
\baselinereflection\ model, and systematic residuals persist in the \chandra\ data between
$\sim$1--5\,\kev. Although the best fit absorber for the \chandra\ data remains Compton-thin (\nh\ =
$8.40^{+1.14}_{-0.97} \times 10^{22}\pcmsq$), a similar covering fraction is inferred ($cf =
0.31\pm0.03$). Given the known size of the X-ray emitting region ($\sim$10\,\rg), determined
independently by X-ray microlensing\cite{DaiKochanek2010quasar}, such small covering fractions
require a highly fine-tuned and \textit{ad-hoc} geometry for the absorbing cloud. Furthermore, we
stress that a scenario in which the spectral features are formed by absorprion/reprocessing by
large, distant, distributed structures is also strongly at odds with the known size of the X-ray
source.\\

\vspace*{1cm}
\setcounter{secondbib}{41}

\newpage
\setcounter{figure}{0}

\begin{center}
{\Large Supplementary Figures}
\end{center}

\begin{figure}[ht]
\begin{center}\label{fig1 Supplementary}
\includegraphics[width=0.48\textwidth]{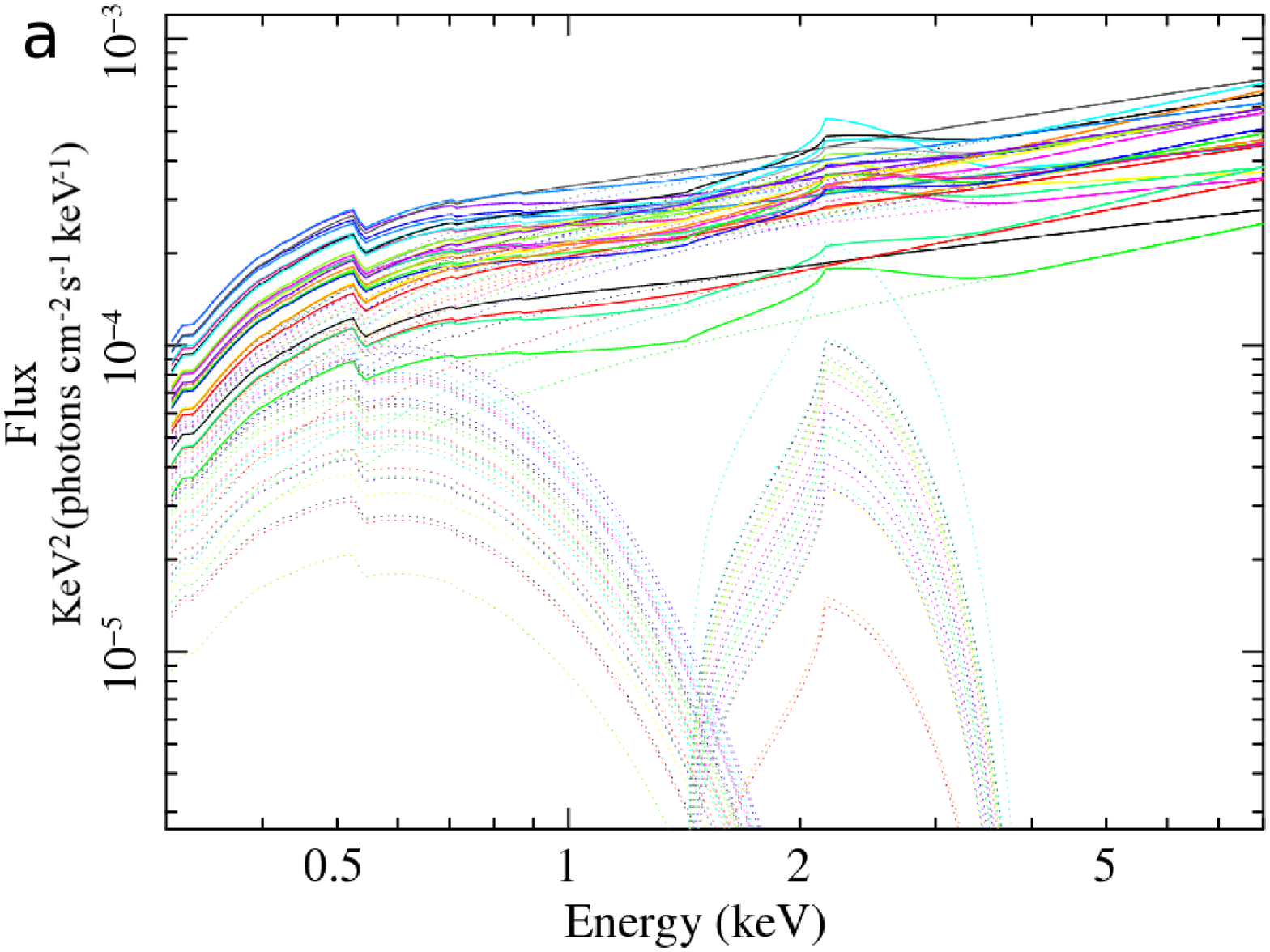}
\includegraphics[width=0.48\textwidth]{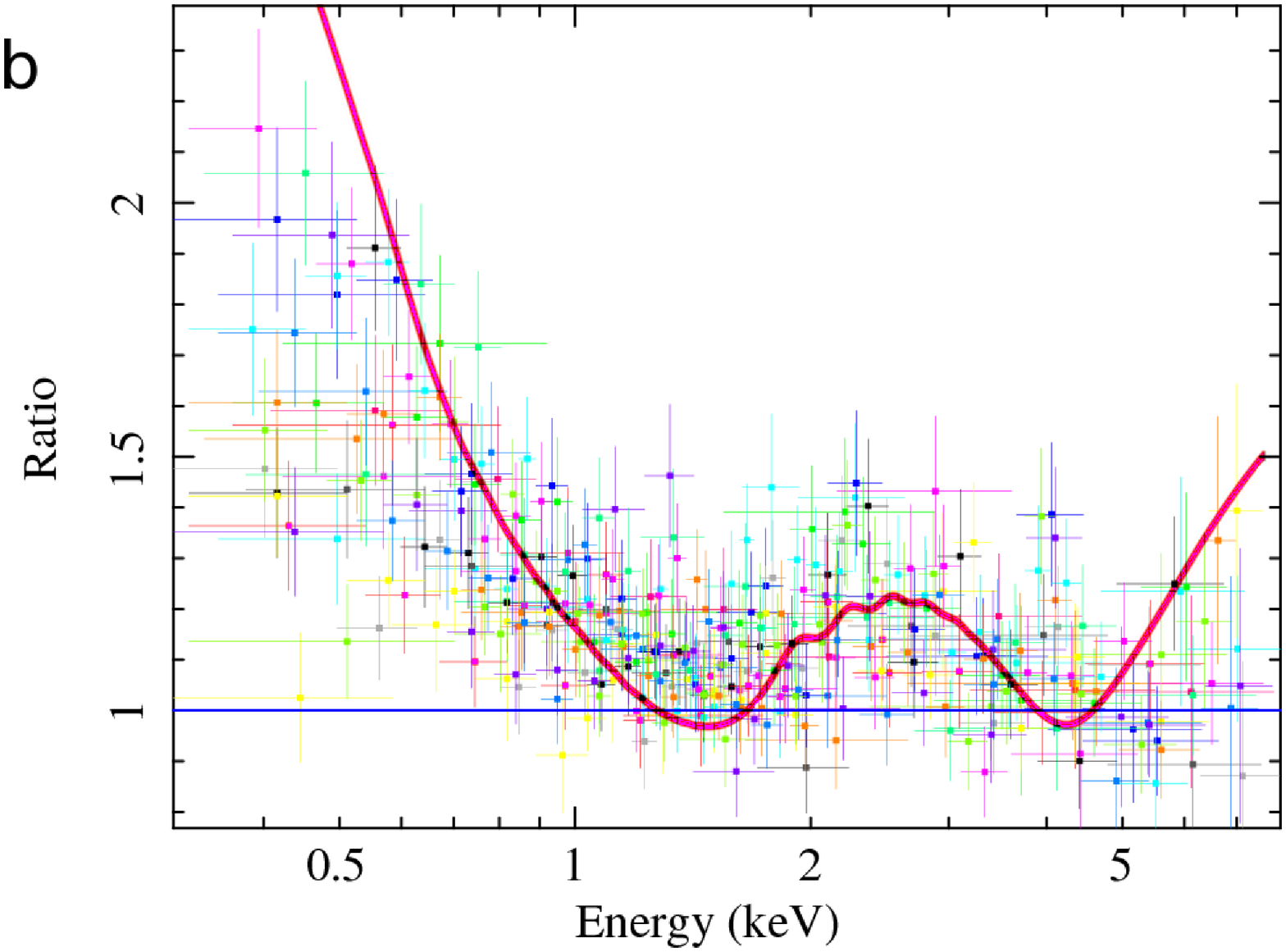}
\caption{ \textbf{a:} Phenomenological models for all 27 \chandra\ observations of Image~B fit with
  a \diskbb\ component to mimic the soft excess as well as a \relline\ line profile to account for
  the residuals in the 2-4\kev\ range and a power-law for the continuum. The normalisation of the
  various components as well as as the power-law indices are allowed to vary between epochs,
  however, the intrinsic column density, the ionisation state (related to the centroid of the line),
  emissivity profile and inclination of the disk are assumed to be unchanging between the various
  epochs.  {\bf b:} Ratio plot made in a similar manner as in {Extended Data Fig. 2} for all 27
  epochs. The solid magenta line is the baseline theoretical best fit model-to-power-law ratio
  expected from the self-consistent \reflionx\ model found for epoch~23. The data have been rebinned
  for visual clarity.}
\end{center}
\end{figure}

\begin{figure}[t]
\begin{center}\label{fig2 Supplementary}
\includegraphics[width=0.48\textwidth]{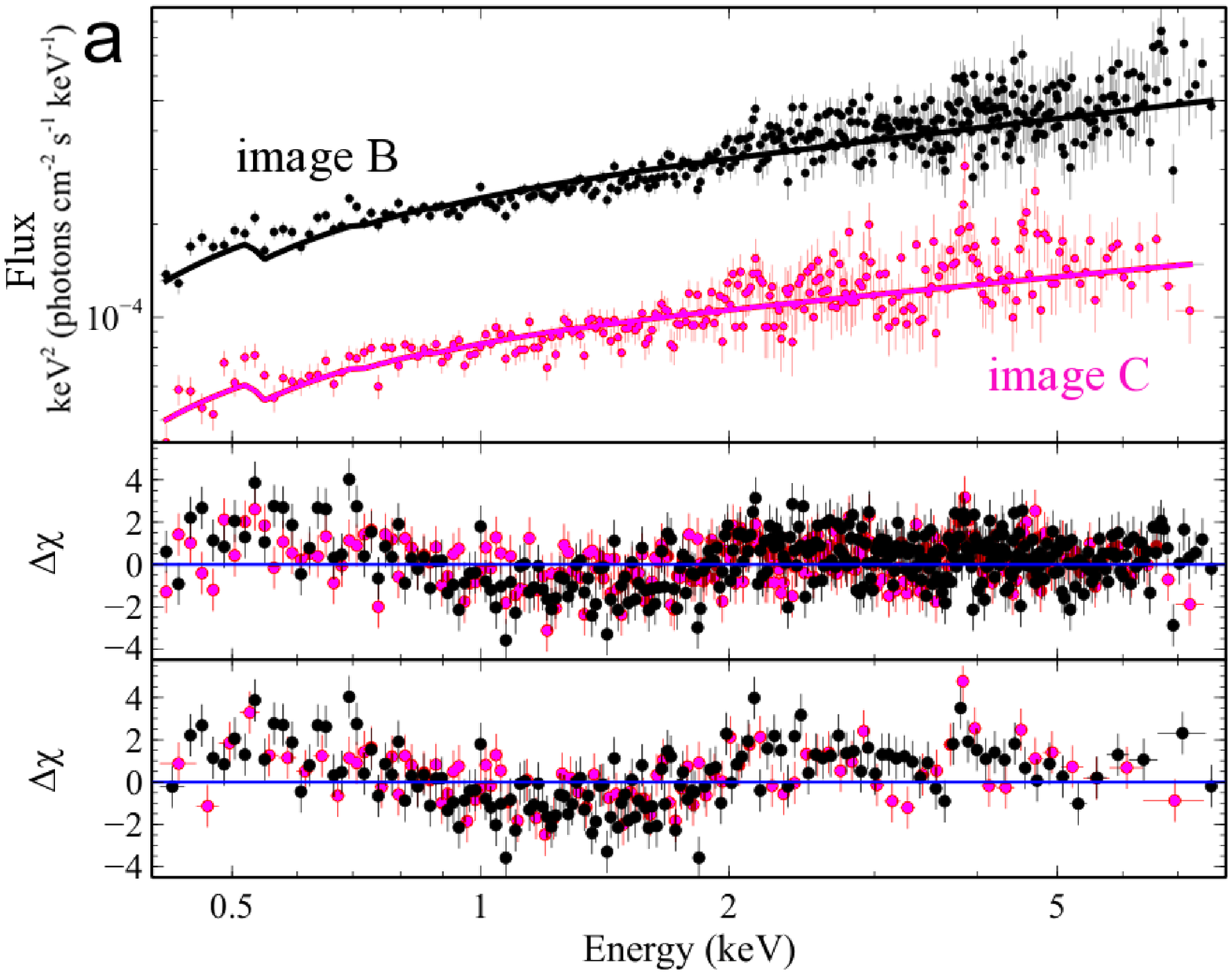}
\includegraphics[width=0.48\textwidth]{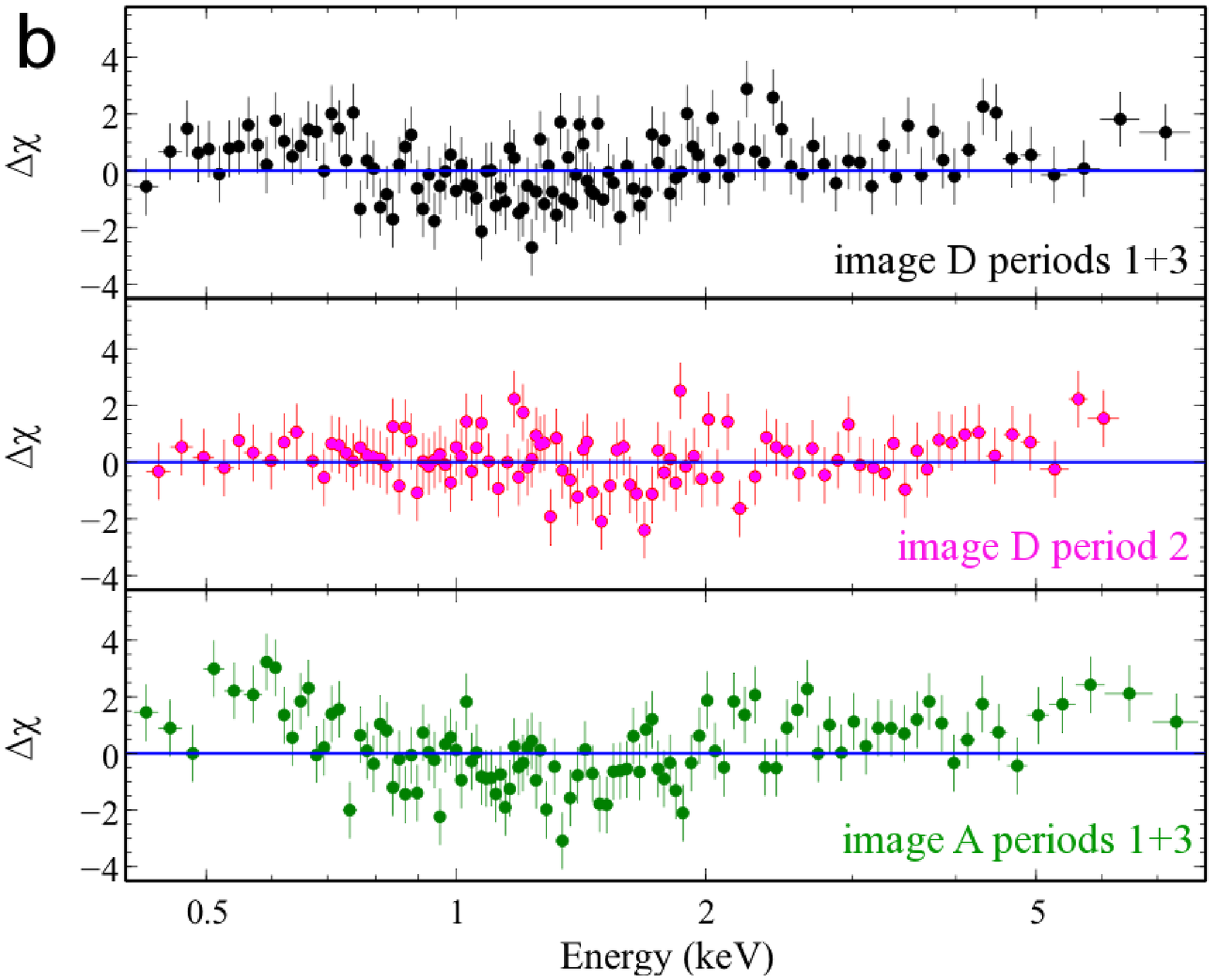}
\caption{{\bf a:} Co-added \chandra\ spectra of Images B (black) and C (magenta) during all
  observations. The spectra are shown fit with a simple power-law, and the residuals to this model
  are shown in units of 1$\sigma$ deviations in the middle panel. The data in the bottom panel have
  been rebinned for display purposes. (\textbf{b:}) Co-added spectra of Periods 1 and 3 (black) and
  Period 2 (magenta) of Image-D shown as residuals to a simple power-law together with Gaussians
  similar to that shown in ref\cite{ChartasKochanek2012quasar}.  The residuals to this model are
  shown in units of 1$\sigma$ deviations. The bottom panel shows similar residuals for the
  corresponding Periods 1+3 of Image-A. The data have been rebinned for display purposes.}
\end{center}
\end{figure}

\begin{figure}[t]
\begin{center}\label{fig3 Supplementary}
\includegraphics[width=0.48\textwidth]{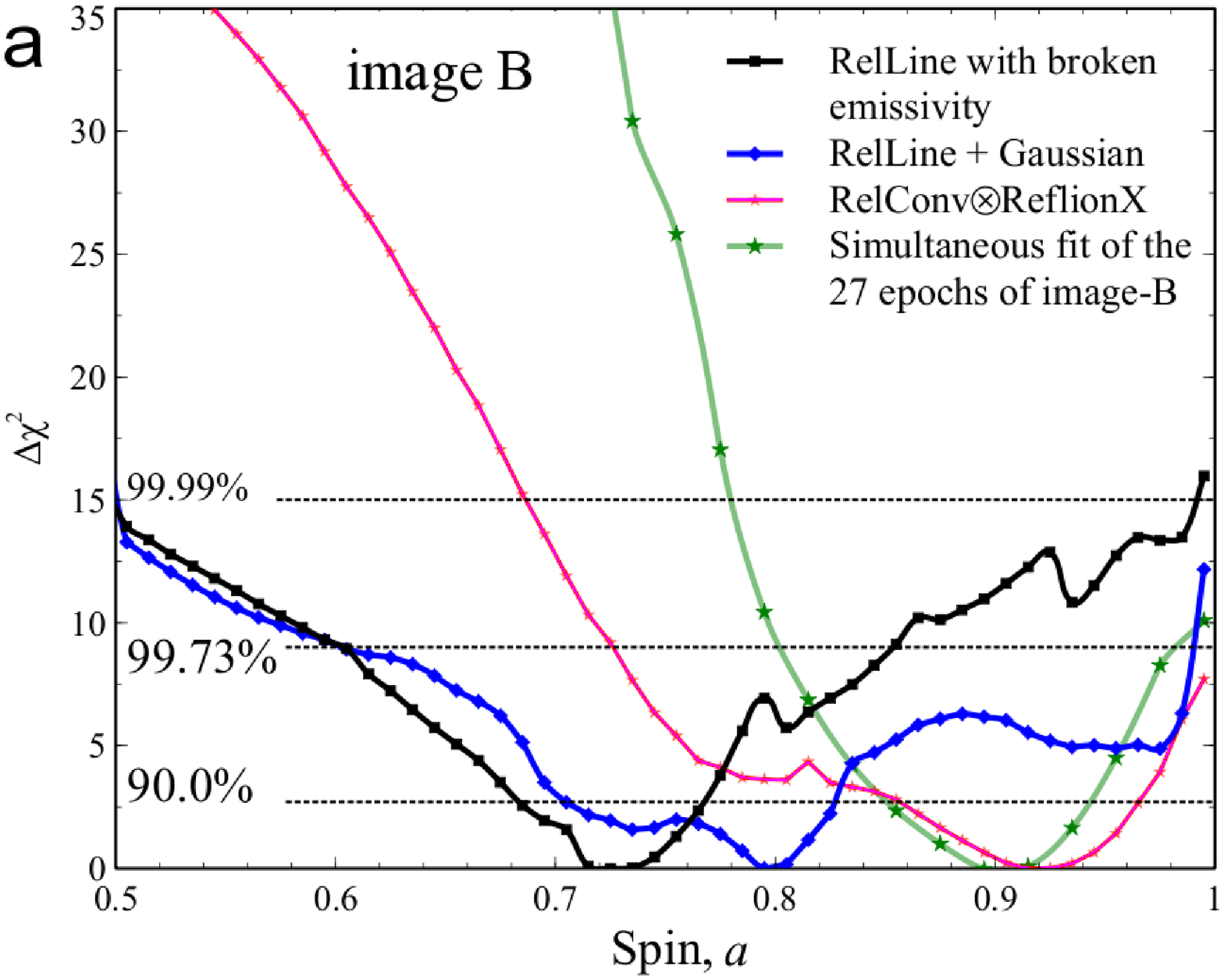}
\includegraphics[width=0.48\textwidth]{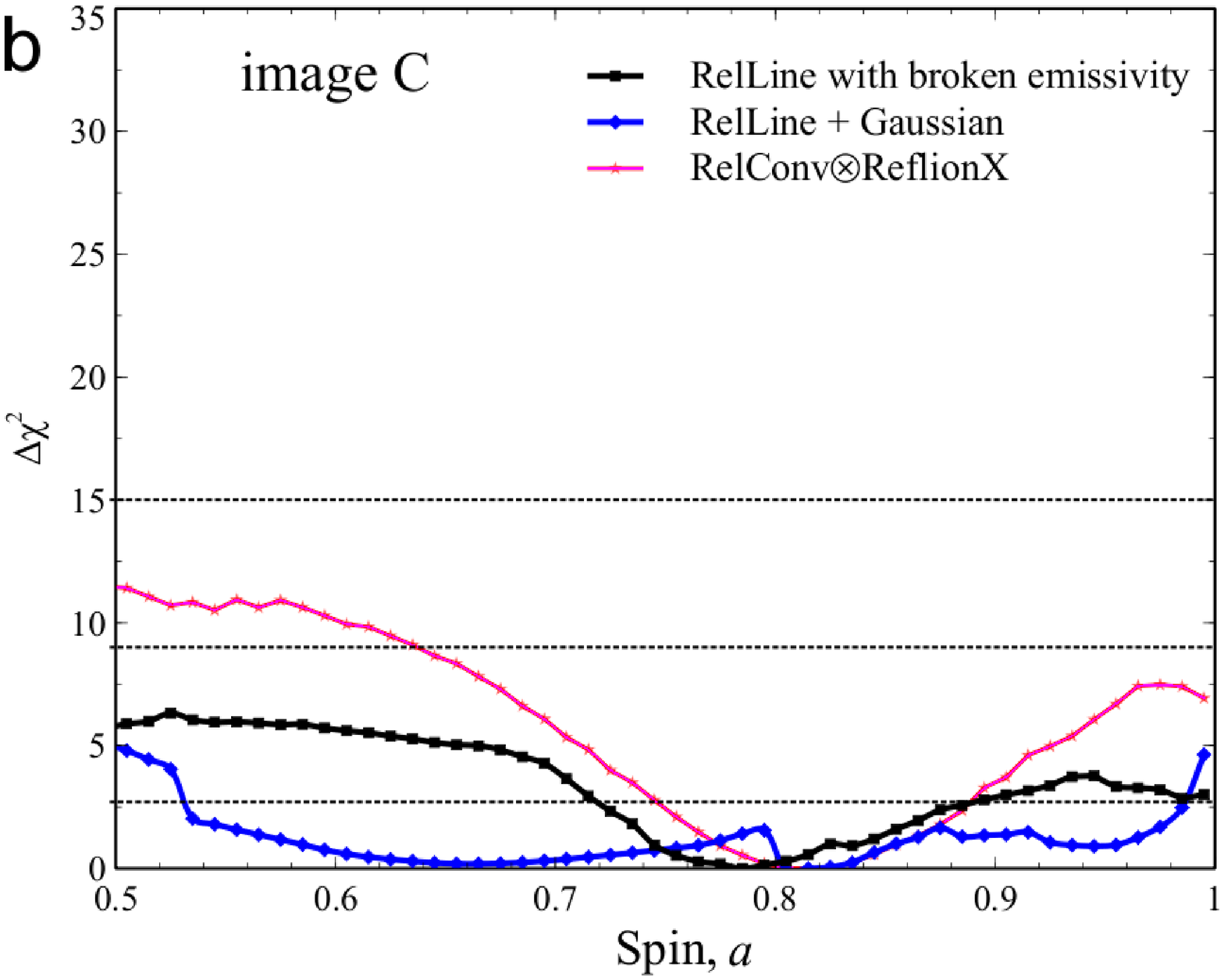}
\includegraphics[width=0.48\textwidth]{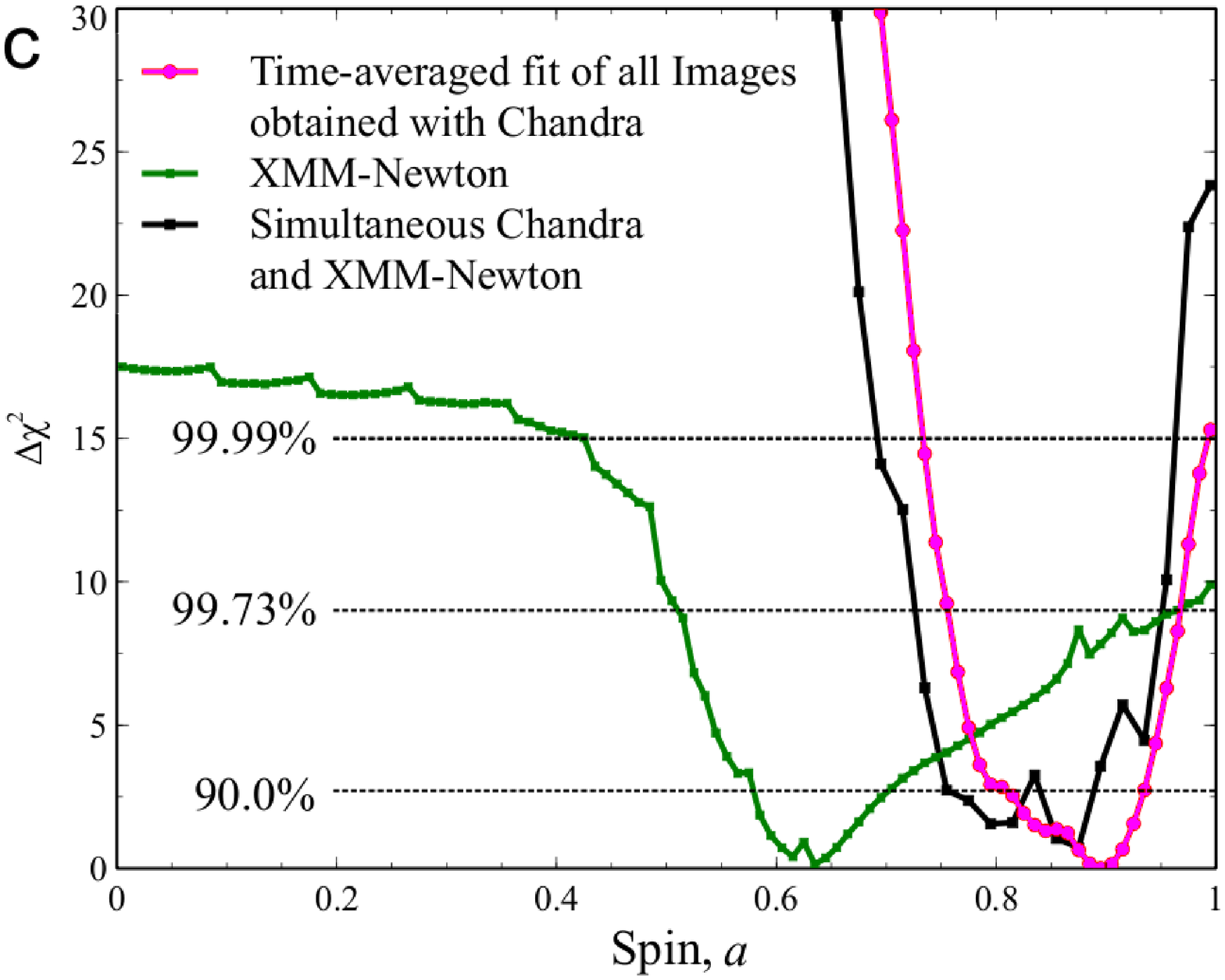}
\caption{ Goodness-of-fit versus spin parameter for the co-added spectrum of Images-B (\textbf{panel
    a}) and C (\textbf{panel b}). By co-adding the spectra we are probing the time averaged
  behaviour of the reflection spectrum.  Fits were made with the spin parameter varying from 0.495
  to 0.995 in steps of 0.01. The blue contours were made using the relativistic line model
  \relline\ assuming a broken emissivity profile and a further narrow Gaussian line at 6.4\kev. The
  black contours assumes a broken emissivity without an extra narrow Gaussian line. Finally, the
  magenta contours are for the self-consistent \reflionx\ model together with the \relconv\ blurring
  kernel and a Gaussian line at 6.4\kev\ (see text for details). We have also included for Image-B,
  the contour (green) found for the time-resolved analyses described in \S6. \textbf{Panel c:}
  Co-added spectrum of all \chandra\ (magenta contours), \xmm\ (green contours) and simultaneous
  data (black contour). These contours are made with the \baselinereflection\ model. The dotted
  lines show the 99.99\%, 99.73\% ($3\sigma$) and 90\% confidence limit.}
\end{center}
\end{figure}

\begin{figure}[t]
\begin{center}\label{fig4 Supplementary}
\includegraphics[width=0.48\textwidth]{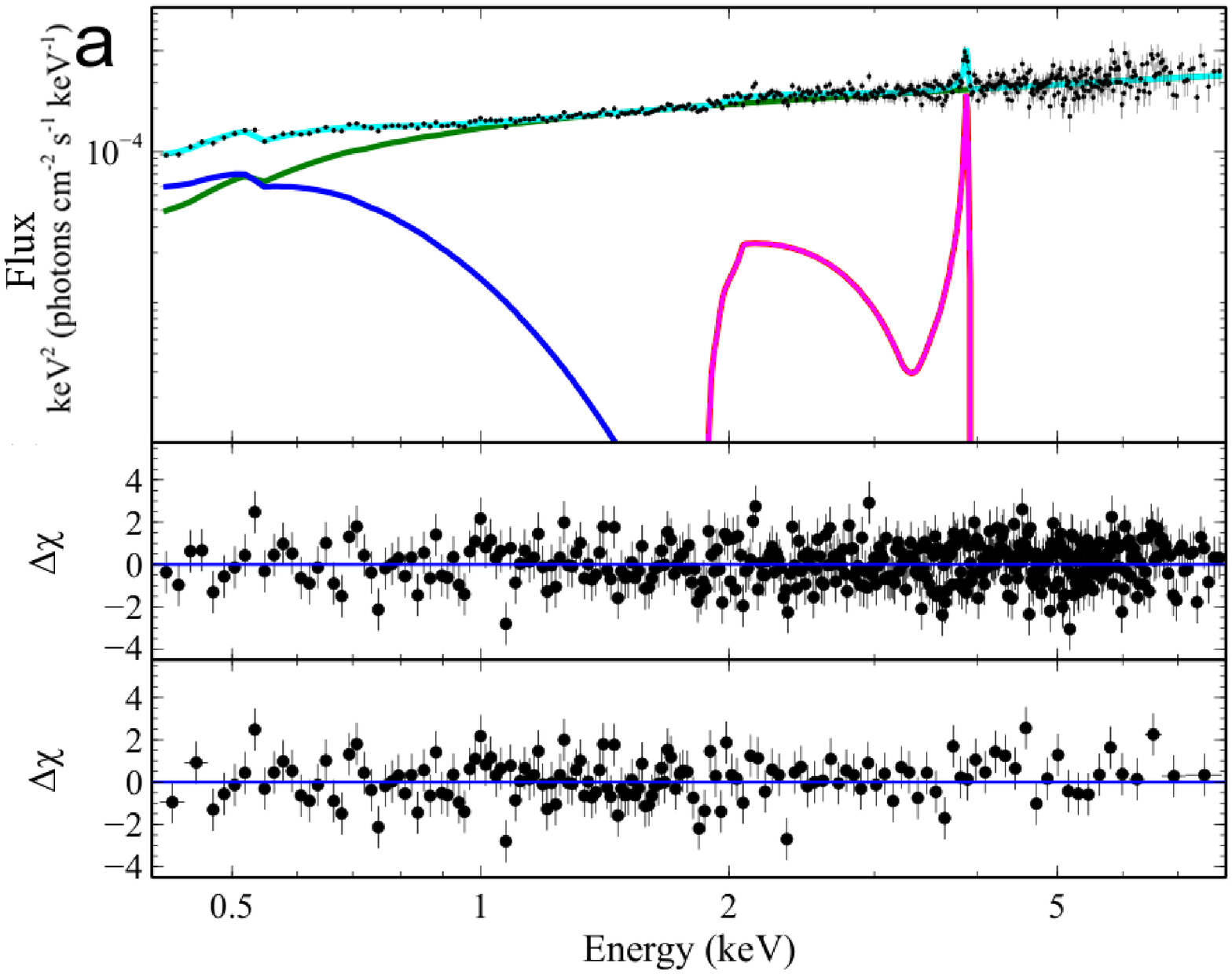}
\includegraphics[width=0.48\textwidth]{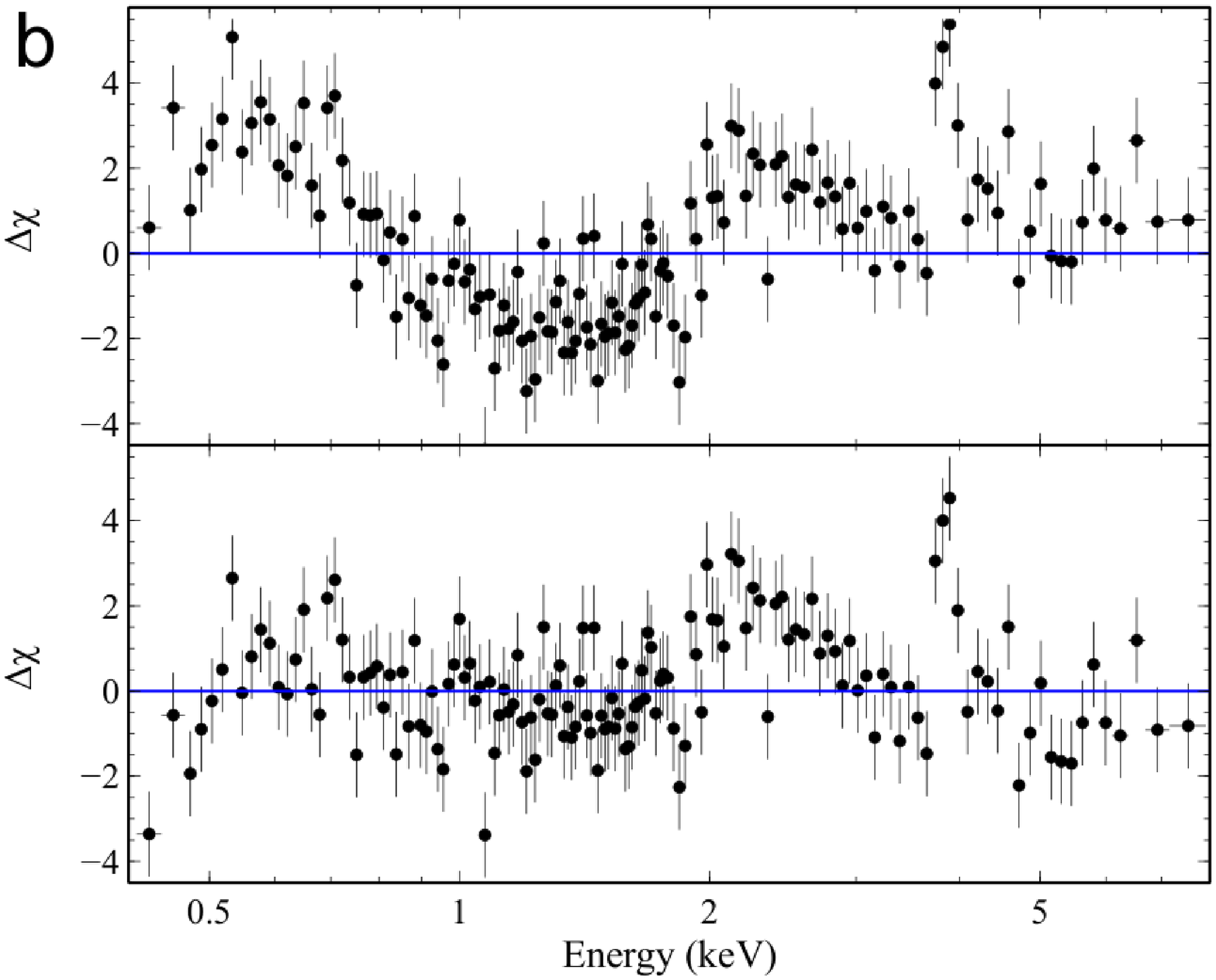}
\caption{ {\bf a:} Co-added spectra of all \chandra\ images. The unfolded spectrum is shown to the
  best fit simple model with the relativistic line, soft disk component and power-law shown in
  magenta, blue and green respectively. The residuals to this model are shown in units of 1$\sigma$
  deviations in the middle panel. The bottom panel shows the residuals after rebinned the data for
  display purposes. \textbf{b:} Top panel shows the residuals to a simple power-law and the bottom
  panel shows similar residuals after the addition of a \diskbb\ component.}
\end{center}
\end{figure}

\begin{figure}[ht]
\begin{center}\label{fig5 Supplementary}
\includegraphics[width=0.48\textwidth]{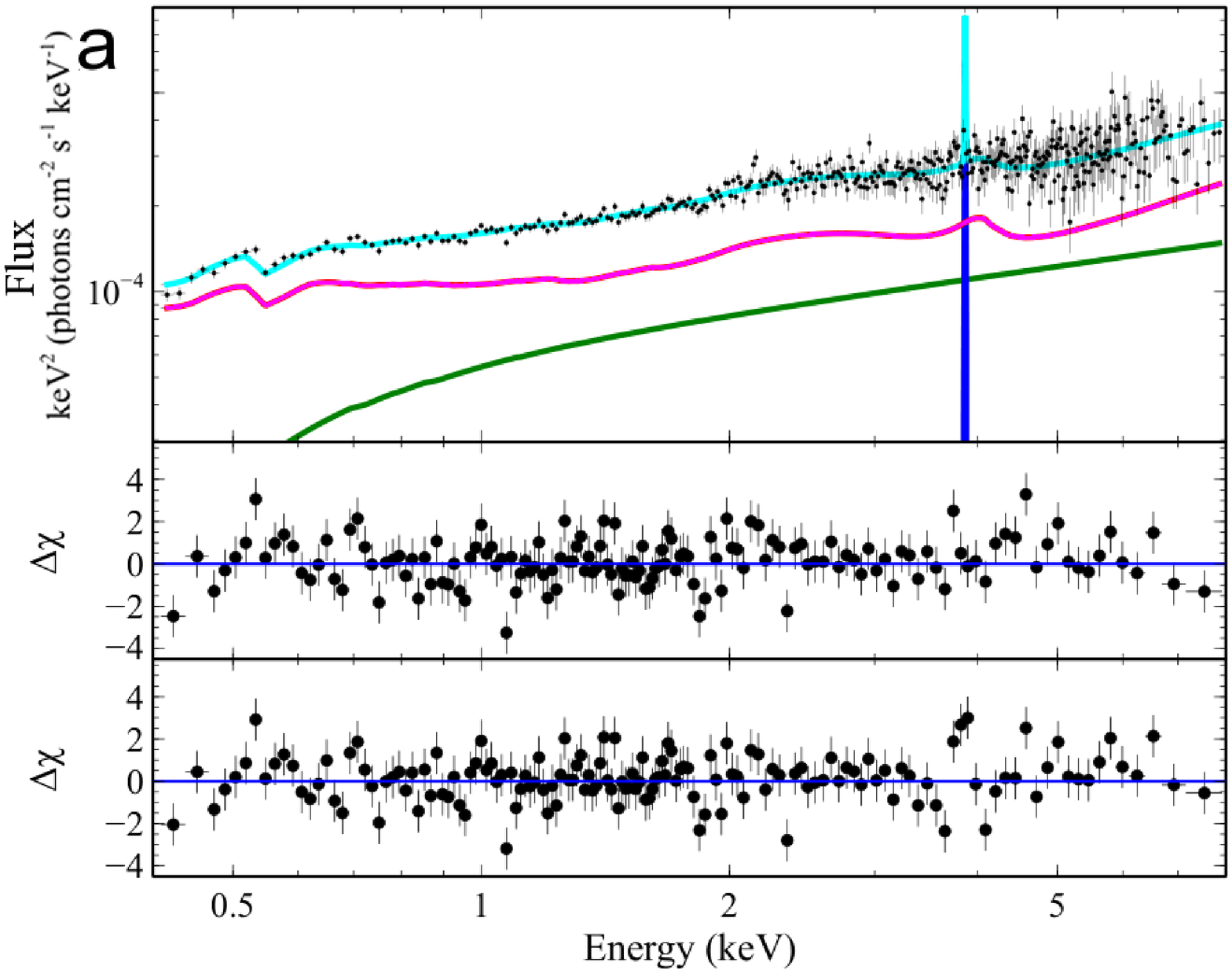}
\includegraphics[width=0.48\textwidth]{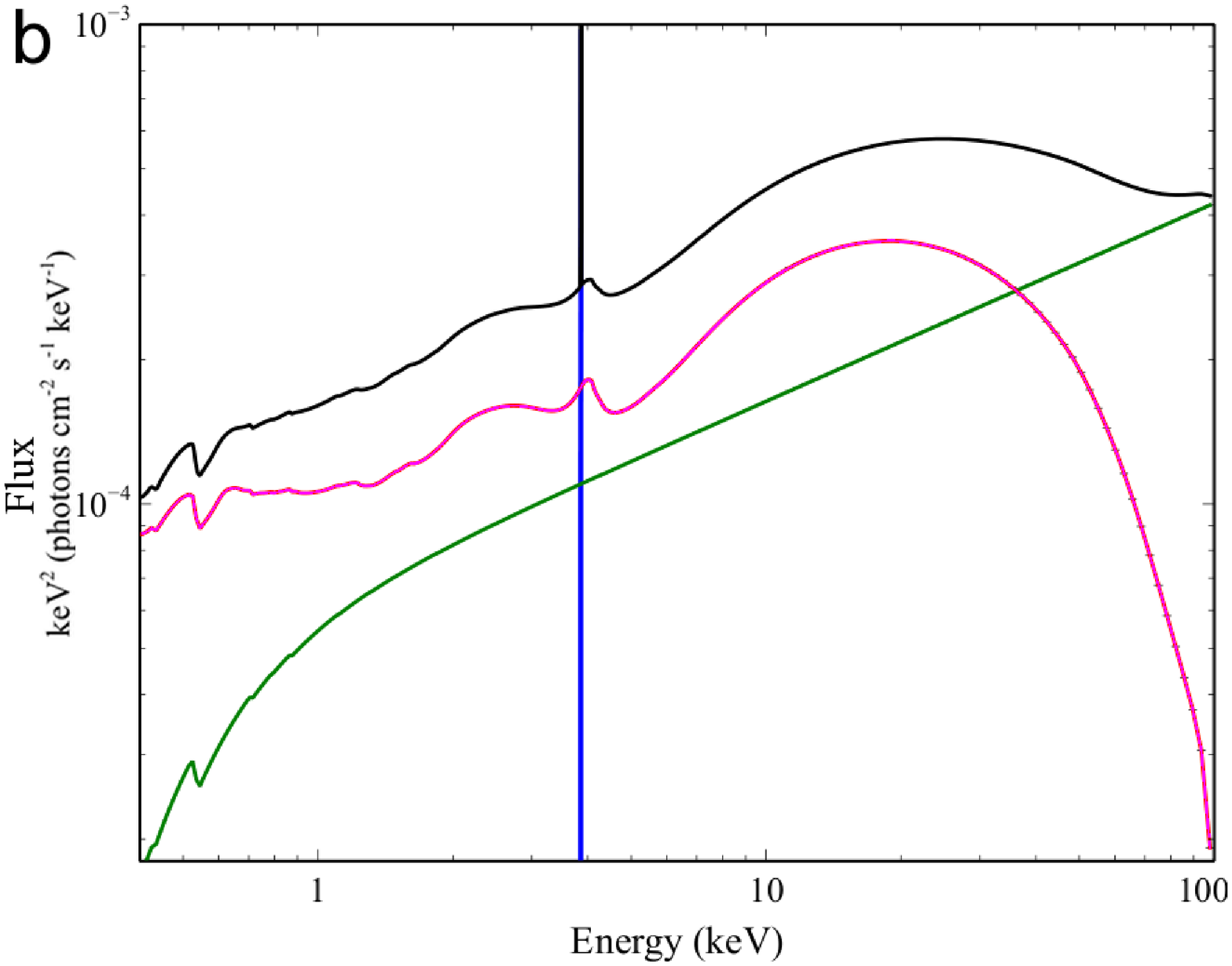}
\caption{ {\bf a:} Co-added spectrum of all \chandra\ epochs.  The spectrum was fit with a
  physically motivated, self consistent model for both the soft-excess and the broad iron line.  The
  residuals to the model is shown in the middle panel. The \reflionx, power-law and narrow Gaussian
  components are shown in magenta green and blue respectively, with the total model shown in
  cyan. The bottom panel shows the residuals without the narrow Gaussian component. (\textbf{Panel
    b:}) Extrapolated model showing a reflection dominated continuum.  The data was rebinned for
  display purposes.}
\end{center}
\end{figure}

\begin{figure}[ht]
\begin{center}\label{fig6 Supplementary}
\includegraphics[width=0.48\textwidth]{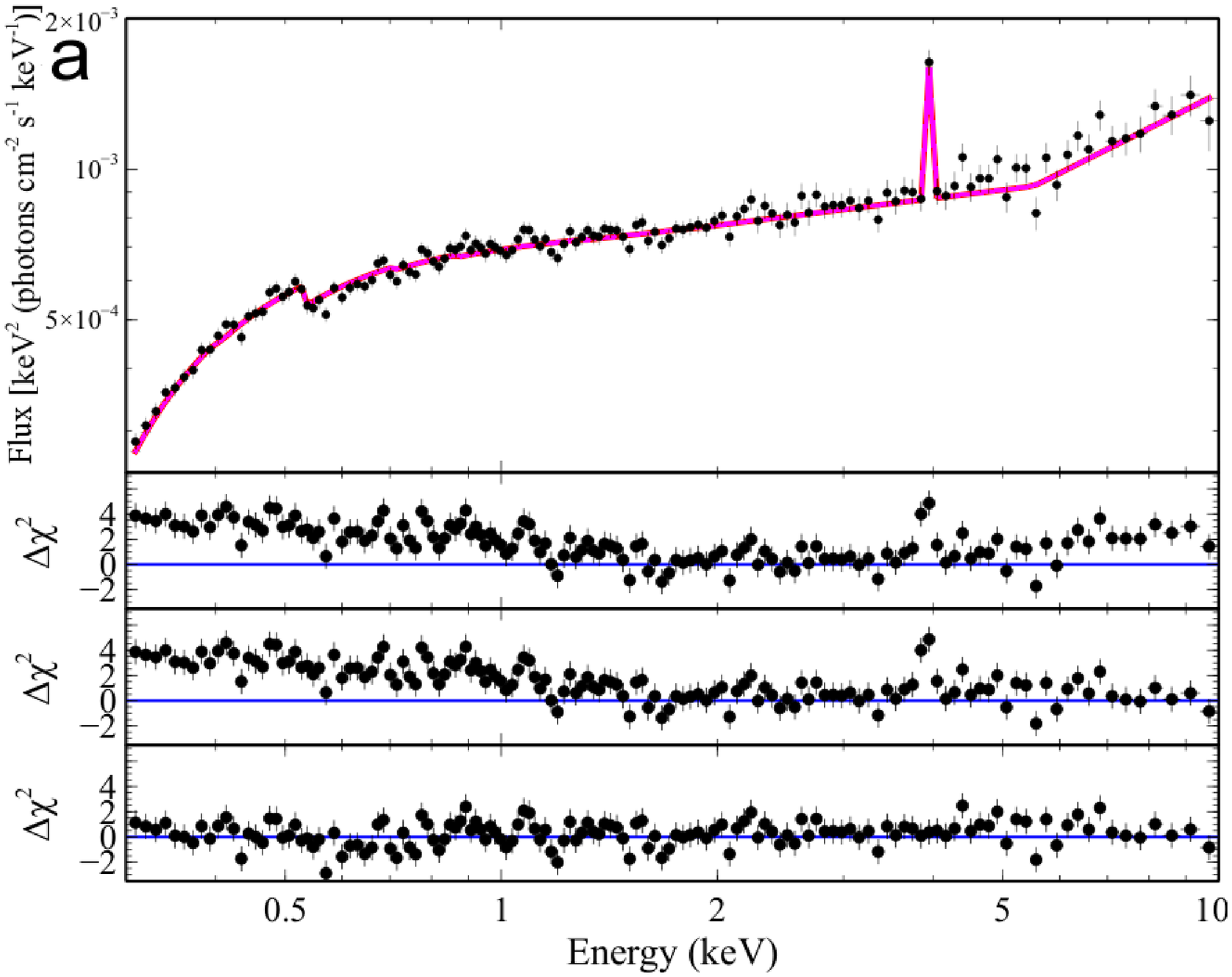}
\includegraphics[width=0.48\textwidth]{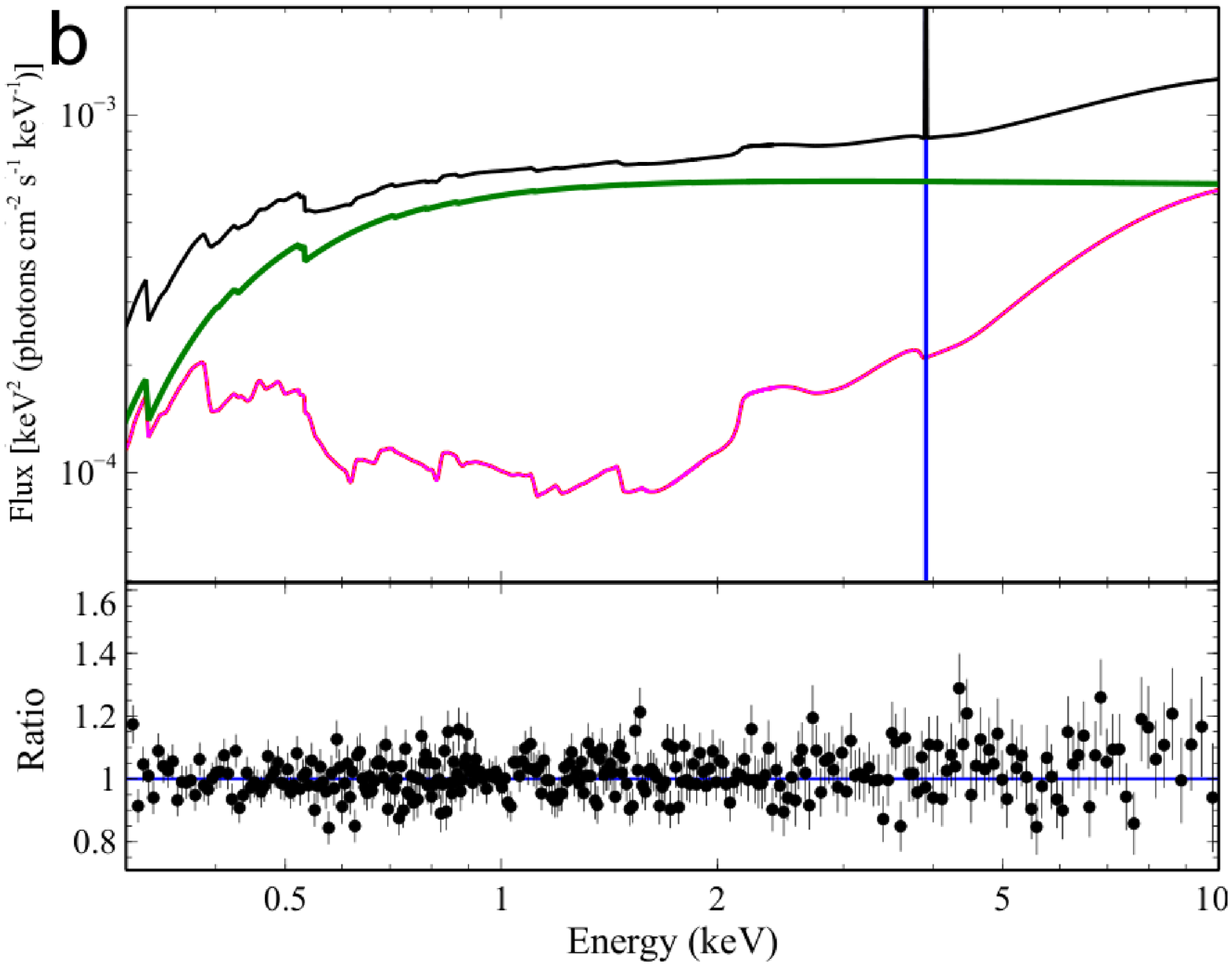}
\caption{ {\bf a:} \xmm\ \epicpn\ spectrum of \rx. (Top) unfolded spectrum showing the total broken
  powerlaw together with a soft excess and narrow Gaussian in magenta. The second panel from the top
  shows the residuals upon removal of the soft component, the narrow line and break in the powerlaw.
  The following panel shows the fit with the addition of a break in the powerlaw and the bottom
  panel shows the best phenomenological fit to the \xmm\ data.  {\bf b:} \xmm\ data now fit with the
  \baselinereflection\ model. The reflection component is again shown in magenta, with the total,
  powerlaw and narrow line shown in black, green and blue respectively. The bottom panel shows the
  ratio to this model. The data have been rebinned for display purposes.}
\end{center}
\end{figure}

\begin{figure}[ht]
\begin{center}\label{fig7 Supplementary}
\includegraphics[width=0.9\textwidth]{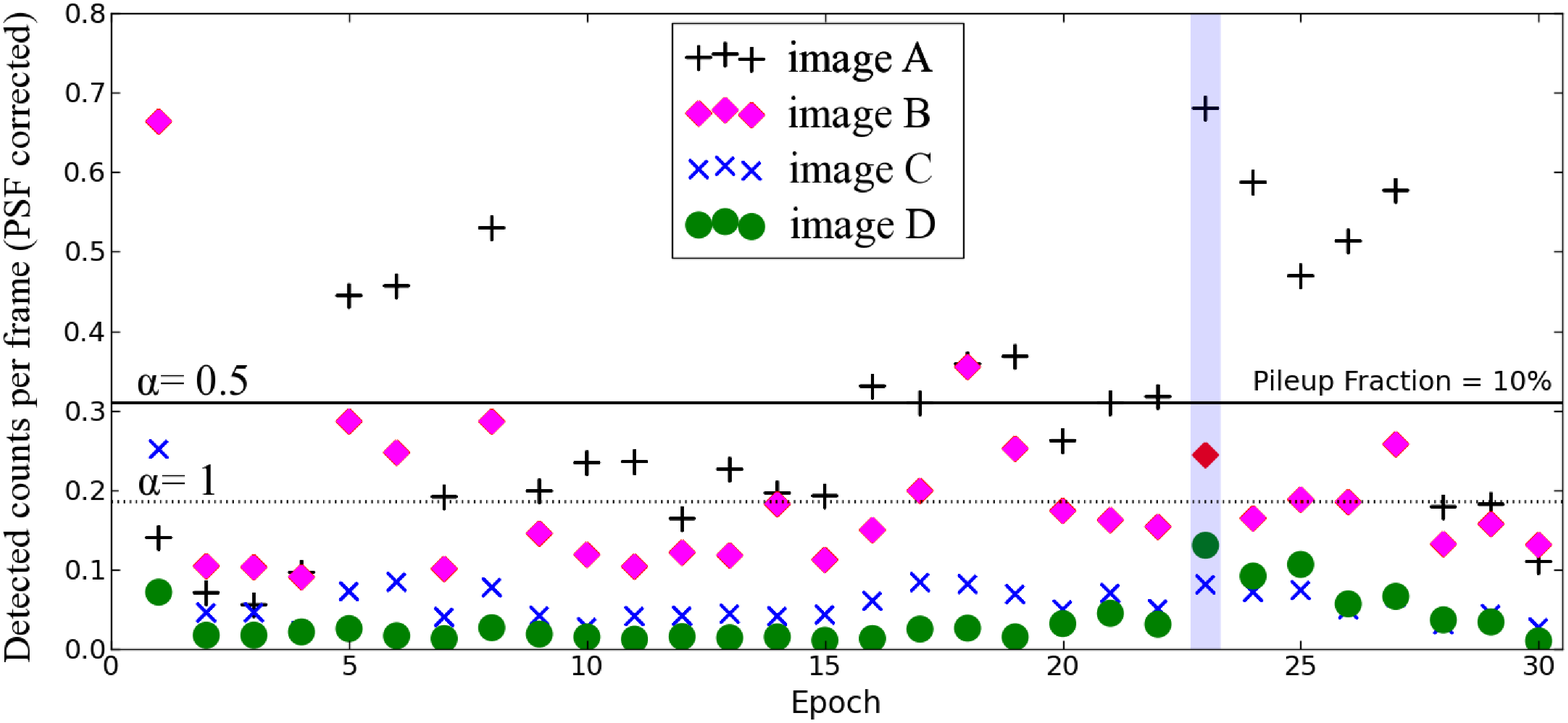}
\caption{ Detected Counts per Frame as a function of observation epoch. The counts have been
  PSF-corrected in order to account for the size of the extraction region.  As such, the counts
  presented are approximately a factor of 1.69, 1.71, 1.69 and 1.21 higher that that observed for
  images~A, B, C and D respectively.  We show with solid and dotted horizontal lines the expected
  10\% pileup fraction assuming a grade migration parameter of $\alpha=1$ and 0.5, respectively. The
  highlighted epoch corresponds to that used in the in-depth study of possible spectral distortions
  due to pile-up/cross contamination.  We use this epoch to assess the implication of high count
  rate on our results as it presents the greatest chance of cross-contamination due to the peak
  brightness of Image-A. The work highlighted in the Online methods clearly shows that the features
  observed in this epoch is consistent with those observed in all, fainter, observations.}
\end{center}
\end{figure}

\begin{figure}[ht]
\begin{center}\label{fig8 Supplementary}
\includegraphics[width=0.48\textwidth]{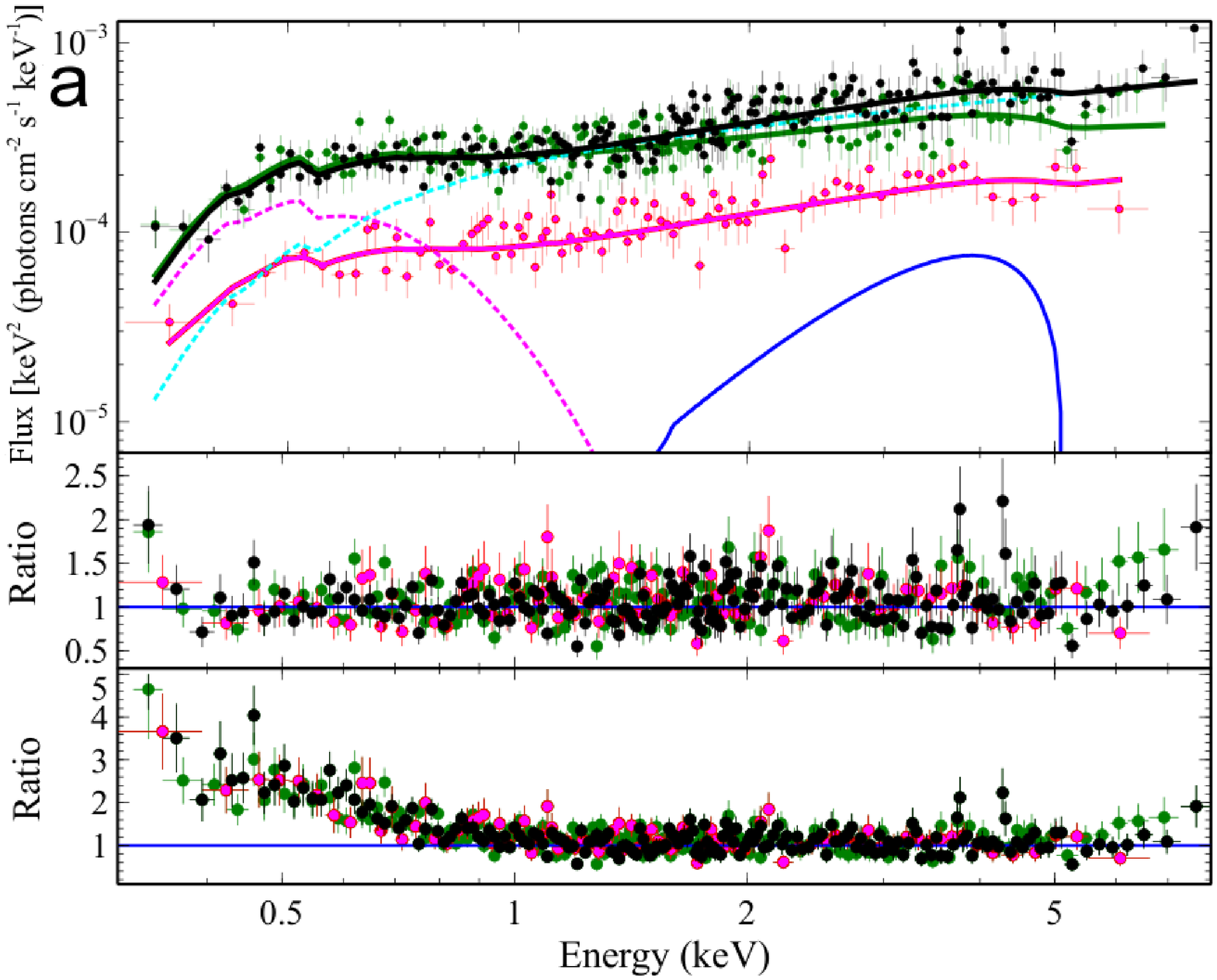}
\includegraphics[width=0.48\textwidth]{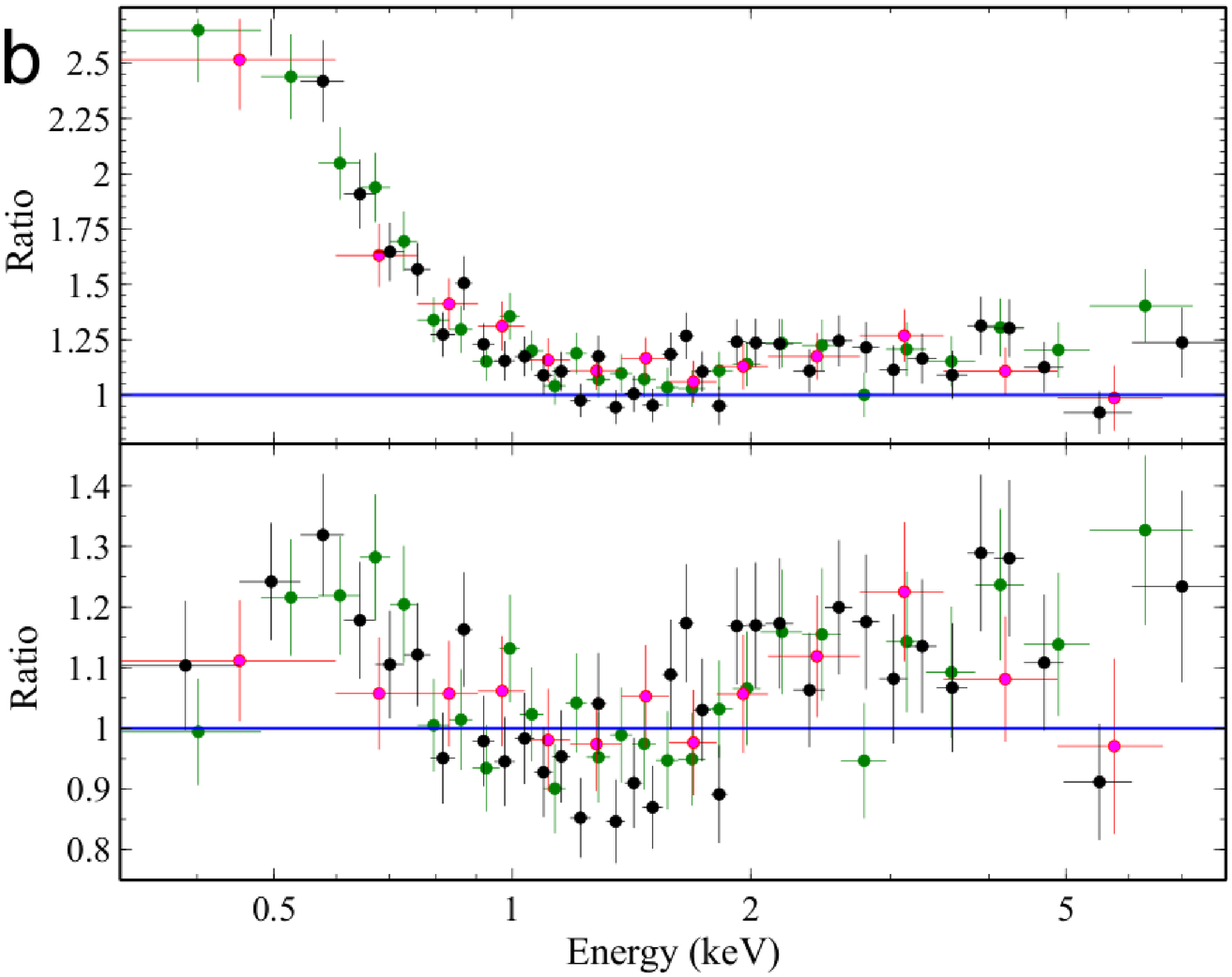}
\caption{{\bf a:} \chandra\ Spectra of images-B, C and D from the observation made on 2009 November
  28. This observation is highlighted in Extended Data Fig.~1 and is representative of the brightest
  \chandra\ epochs used in this work. It is also the observation in which Image-A is at its
  brightest and therefore presents the largest chance of cross contamination as sources are only
  $\sim 1$\as\ apart. The spectra were fit phenomenologically with a combination of a power-law,
  \diskbb\ and \relline\ line profile to describe the continuum, soft excess and iron emission
  feature respectively. The ratio to the model are shown both with (centre) and without (bottom) the
  disk component representing the soft-excess. The spectra of image-B, C and D are shown in black,
  magenta and green respectively. The power-law, \diskbb\ and \relline\ components for image-B are
  shown in cyan, magenta, and blue respectively with the total model shown as solid lines for the 3
  images. {\bf b:} Data-to-model ratio obtained after removing the \relline\ and \diskbb\ components
  (top) and after refitting with a single power-law (bottom). The ratio for images B, C and D are
  shown in black, magenta and green respectively and have been rebinned for display
  purposes. Despite the different flux levels between these 3 images, it is clear that the residuals
  to a simple powerlaw remains similar, suggesting that pileup is not significant in the brightest
  observations Image-B.}
\end{center}
\end{figure}

\clearpage
\newpage
\begin{center}
{\Large Supplementary Tables}
\end{center}

\vspace{10mm}
\setcounter{table}{1}
\begin{table}[ht]\label{table1}
{\sffamily\noindent\textbf{Table 1}~|\hspace{0.3em}}{Model summary for the co-added spectra of Images-B and C.  }
\begin{center}
\begin{normalsize}
\begin{tabular}{lccccc}
\hline\\[-2.0ex]
    & \multicolumn{2}{c}{{Baseline-simple}}  &  \multicolumn{2}{c}{{Baseline-reflection}}\\[0.5ex] 
   parameters  & image B & image C & image B & image C  \\[0.5ex]
   \hline\\[-2.0ex]
   $N_{\rm H}(z=0.658)$ ($\times 10^{21}\pcmsq$)& $1.4^{+0.8}_{-0.4}$  & $<2$  &$1.1^{+0.3}_{-0.5}$  &$<0.8$ \\[0.5ex]
   $\Gamma$ & $1.65^{+0.03}_{-0.02}$ & $1.71^{+0.06}_{-0.04}$ &   $1.62^{+0.15}_{-0.11}$ & $1.55^{+0.13}_{-0.15}$\\[0.5ex]
   $N_{\rm power-law}$ ($\times10^{-4}$) & $2.6\pm0.1$ & $0.85\pm0.08$ &$<1.19$ • &  $0.38^{+0.29}_{-0.14}$\\[0.5ex]
   $T_{\rm Diskbb}$ (\kev) & $0.12\pm0.01$  & $0.15^{+0.09}_{-0.04}$ & •  --- • &---\\[0.5ex]
   $N_{\rm Diskbb}$ & $200^{+270}_{-140}$  &  $14^{+90}_{-9}$  & •  --- • &---\\[0.5ex]
   $E_{\rm RelLine}$  (\kev) & $6.43^{+0.05} $ & $6.49\pm0.06$  & •  --- • &---\\[0.5ex]
   $Q_{\rm in}$ & $7.8^{+0.7}_{-1.9} $ &$6.5^{+0.5}_{-1.1} $ & •  $6.1^{+0.4}_{-0.7} $  &$7.0^{+0.9}_{-1.8} $\\[0.5ex]
   $Q_{\rm out}$  & $<2.7$ & $<2.3$ &    $< 3.7$ & $< 8$\\[0.5ex]
   Inclination & $<13$ &$<16$  &  $26^{+8}_{-14}$ & $<27$\\[0.5ex]
   $N_{\rm RelLine}$ ($\times10^{-6}$) & $7.0^{+2.5}_{-1.7}$  &$3.3.0^{+1.6}_{-1.2}$  &• --- • &---\\[0.5ex]
   $N_{\rm ReflionX}$  ($\times10^{-8}$) & ---  & --- & $ 6.9^{+3.3}_{-3.5}$  &$ 0.45^{+0.05}_{-0.03}$\\[0.5ex] 
   $\xi$  {\rm ($\ergcmps$)}& --- & ---  &  $623^{+230}_{-210}$ & $1290^{+890}_{-870}$\\[0.5ex] 
   ${\rm f_{\rm reflect}/f_{\rm illum}}$ $(0.1-10)$\kev& --- & ---& $3.6\pm2.5$  & $1.4\pm1.2$\\[0.5ex]
   ${\rm f_{\rm reflect}/f_{\rm illum}}$ $(0.1-100)$\kev& --- & ---& $2.6\pm2.0$  & $0.9\pm0.6$\\[0.5ex]
   Spin ($a= {\rm Jc/GM^2}$)  &$0.73\pm0.04$ &$0.79^{+0.11}_{-0.07}$  &$0.92^{+0.04}_{-0.06}$ &$0.80^{+0.08}_{-0.06}$\\[0.5ex]
   $\chisq/\nu$ & $393.9/350$ &$ 231.0/242$& •$ 407.5/351 $ & $238.6/243$\\[0.5ex] 
\hline 
\end{tabular} 
\end{normalsize}
\end{center}
\begin{footnotesize}
\vspace*{1mm}
\textbf{Notes:}
$^{a}$ Rest frame energy constrained between 6.4-6.97\kev. $^b$ Value for the outer emissivity constrained to be $>2$ with a break radius frozen at 10\rg. All models include a fixed  Galactic absorption of $N_{\rm H}(z=0)=0.36\times 10^{21}\pcmsq$. $^c$ Reflection fraction defined as the ratio of the reflected to the illuminating continuum in the given energy band. All errors are 90\% confidence on one parameter 
\end{footnotesize} 
\end{table}

\newpage
\begin{table}[ht]\label{table2}
 {\sffamily\noindent\textbf{Table 2}~|\hspace{0.3em}} Model summary for
 the time-averaged \chandra\ spectrum of \rx\ as well as the \xmm\ \epicpn\ data using
 \textit{Baseline-reflection}.
\begin{center}
\begin{normalsize}
\begin{tabular}{lcccc}
\hline\\[-2.0ex]
    &  \multicolumn{2}{c}{Individual}& \multicolumn{2}{c}{Joint}\\[0.5ex]
   parameter& Chandra\ & XMM-Newton  & Chandra\ & XMM-Newton \\ [0.5ex]
   \hline\\[-2.0ex]
    $N_{\rm H} (z=0.658$)  ($\times 10^{21}\pcmsq$)&$0.8^{+0.3}_{-0.5}$ &$1.1^{+0.3}_{-0.4}$&\multicolumn{2}{c}{$1.07^{+0.26}_{-0.16}$ }\\[0.5ex]
    $\Gamma$ & $1.59\pm0.08$&$2.02\pm0.04$ & $1.64^{+0.06}_{-0.05}$&$2.02^{+0.04}_{-0.03}$ \\[0.5ex]
    $N_{\rm power-law}$  ($\times10^{-5}$) & $6.3^{+2.5}_{-2.1}$ & $187^{+5}_{-4}$& $9.4^{+2.9}_{-2.3}$ & $187\pm4$\\[0.5ex]
    $Q_{\rm in}$ & $6.0^{+0.8}_{-0.4}$ & $>5.4$ & $>5.9$ & $5.0^{+1.4}_{-1.5}$\\[0.5ex]
    $Q_{\rm out}$  & $<2.4$ & $3.3^{+1.1}_{-0.7}$& $2.9^{+0.2}_{-0.8}$ & $3.3^{+1.0}_{-1.3}$\\[0.5ex]
   $R_{\rm break}$ ($r_{\rm G} = {\rm GM/c^2}$) &10(f) & $6^{+4}_{-1}$&$5.9^{+3.0}_{-1.7}$& $<20$\\[0.5ex]   
     Inclination &  $ 20^{+6}_{-7}$ & $<17$& \multicolumn{2}{c}{$ 15^{+9}_{-15}$} \\[0.5ex]
     $N_{\rm ReflionX}$ ($\times10^{-8}$) & $3.0^{+1.2}_{-1.0}$ & $75\pm34$ & $2.8^{+0.6}_{-1.3}$ & $78^{+44}_{-16}$  \\[0.5ex]
     $\xi (\rm \ergcmps)$ &$725^{+286}_{-118}$& $99^{+15}_{-37}$ &$612^{+195}_{-92}$& $98^{+16}_{-4}$\\[0.5ex]
    ${\rm f_{\rm reflect}/f_{\rm illum}}$ $(0.1-10)$\kev  & $2.3\pm1.2$  & $0.47\pm0.15$ \\[0.5ex]
   ${\rm f_{\rm reflect}/f_{\rm illum}}$ $(0.1-100)$\kev & $1.65\pm0.95$  & $0.53\pm0.13$  \\[0.5ex] 
     Spin ($a={\rm Jc/GM^2}$) & $0.90^{+0.07}_{-0.15}$  & $0.64^{+0.33}_{-0.14}$ &\multicolumn{2}{c}{$a=0.87^{+0.08}_{-0.15} $}   \\[0.5ex]
     $\chisq/\nu$ & $439.6/409$ &$ 849.4/849$ &\multicolumn{2}{c}{$1296.1/1260$ }\\[0.5ex] 
    \hline
\end{tabular}   
\end{normalsize}
\end{center}
\begin{footnotesize}
\vspace{1mm}
\textbf{Notes:} $^a$ Value for the outer emissivity constrained to be $>2$. (f) denotes frozen
values. $^b$ Reflection fraction defined as the ratio of the reflected to the illuminating continuum
in the given energy band. All models include a fixed Galactic absorption of $N_{\rm
  H}(z=0)=0.36\times 10^{21}\pcmsq$. $^c$All errors are 90\% confidence on one parameter except for
the spin where we show the $3\sigma$ limit.
\end{footnotesize} 
\end{table}

\end{document}